\documentclass[aps,physrev,preprint,superscriptaddress,floatfix]{revtex4-2}
\usepackage{amsmath, amsfonts, amssymb, amsthm, mathrsfs, mathtools}
\usepackage{nicematrix} % For nice matrices
\usepackage{booktabs} % For professional looking tables
\usepackage{tabularx}
\usepackage{placeins}
\usepackage{setspace}

\usepackage{siunitx} % use this package module for SI units
\usepackage{enumitem}
\newcommand{\eg}{\textit{e.g.,} }
\newcommand{\ie}{\textit{i.e.,} }
\newcommand{\etc}{\textit{etc.}}
\renewcommand{\S}{Sec.} % Command for section \S
\newcommand{\setb}{II } % Naming the experiment sets. Set 2b is both. 
\newcommand{\setw}{I } % Set 2w is the wave only ones
\usepackage{csquotes}
\renewcommand{\Vec}[1]{\ensuremath{\mathbf{#1}}} % Command for bold vectors instead of arrow vectors
\newcommand{\wnv}{\Vec{k}}  % Wavenumber vector k
\newcommand{\wnxa}[1]{k_{x,#1}}  % Wavenumber vector component k_x
\newcommand{\pwv}{\Vec{k_0}}  % Wavenumber vector k_0, p for primary wave
\newcommand{\pwx}{k_{x, 0}}  % Wavenumber vector component k_x for primary wave
\newcommand{\pwz}{k_{z, 0}}  % Wavenumber vector component k_z for primary wave

\newcommand{\wave}{\mathbb{M}} % Command for referring to wave
\newcommand{\aswamamp}{A_0} % Amplitude of the primary wave from the wavemaker

\usepackage[colorlinks=true, allcolors=blue]{hyperref}
\begin{document}

% Use the \preprint command to place your local institutional report
% number in the upper righthand corner of the title page in preprint mode.
% Multiple \preprint commands are allowed.
% Use the 'preprintnumbers' class option to override journal defaults
% to display numbers if necessary
%\preprint{}

%Title of paper
\title{Experimental observations of sub-threshold triadic resonances in internal wave beams}

% repeat the \author .. \affiliation  etc. as needed
% \email, \thanks, \homepage, \altaffiliation all apply to the current
% author. Explanatory text should go in the []'s, actual e-mail
% address or url should go in the {}'s for \email and \homepage.
% Please use the appropriate macro foreach each type of information

% \affiliation command applies to all authors since the last
% \affiliation command. The \affiliation command should follow the
% other information
% \affiliation can be followed by \email, \homepage, \thanks as well.
\author{Laura J. Irvine}
\email[Corresponding author: ]{l.irvine@imperial.ac.uk}
%\homepage[]{Your web page}
%\thanks{}
\affiliation{Department of Civil and Environmental Engineering, Imperial College London, London, SW7 2AZ, United Kingdom}\affiliation{Department of Applied Mathematics and Theoretical Physics, University of Cambridge, Cambridge, CB3 0WA, UK}

\author{Katherine M. Grayson}
\affiliation{Earth Sciences Department, Barcelona Supercomputing Center, Barcelona 08034, Spain}

\author{Andrew G. W. Lawrie}
\affiliation{Hele-Shaw Laboratory, Queen's Building, University of Bristol, University Walk, Bristol, BS8 1TR, United Kingdom}

\author{Stuart B. Dalziel}
\affiliation{Department of Applied Mathematics and Theoretical Physics, University of Cambridge, Cambridge, CB3 0WA, UK}

\date{13 October 2026}

\singlespacing
\begin{abstract}

This work examines sub-threshold triadic responses to a vortex-ring-induced perturbation of an internal gravity-wave field, using sparsity-promoting dynamic mode decomposition (spDMD) to diagnose wave structures from experimental observations. Compared with a standard approach to dynamic mode decomposition that selects modes by truncation of the singular value spectrum, spDMD instead seeks to minimise reconstruction error and is shown here to robustly identify previously indistinguishable oscillatory signals of high frequency in cases of both triadic resonance instability (TRI) and sub-threshold sustained triadic response (STR). These signals are consistent with wave—wave interaction products composed of a primary internal wave with frequency $\omega_0$, and secondary TRI waves with frequencies $\omega_1$ and $\omega_2$ respectively, leading to newly observed signals with frequencies $\omega_3 = \omega_0 + \omega_1$ and $\omega_4 = \omega_0 + \omega_2$. While signals containing these frequencies have been reported in previous work, including being observed in wave-attractor experiments, their temporal evolution and spatial extent has received much less attention, and we address this in particular here. We show that these signals satisfy conditions for resonance, but fail to satisfy the dispersion relation for internal waves. Most of these high-frequency signals lie above the natural buoyancy frequency $N$, indicating that they are confined, non-propagating oscillatory signals rather than freely propagating internal waves. Despite the transience of their initiation by vortex rings in our experiments, these signals persist throughout the remainder of our experiments, suggesting that they exist due to sustained forcing that is caused by interactions between the primary and secondary internal waves. A hierarchical analysis of wave--wave interactions provides further evidence that feedback between both branches of the TRI secondary waves gives rise to these signals. Whilst we expect our observations to generalise to many classes of internal wave--wave interaction, we cannot exclude the possibility that the particular observations we make in the laboratory may arise due to reflections of the primary wave from a free surface. However, beneath strongly density-stratified regions of the ocean we would expect to find these interaction pathways encountering the same geometric configuration, so we are confident our findings will transfer directly to the natural environment.

\end{abstract}

%\maketitle must follow title, authors, abstract, and keywords
\maketitle

\section{Introduction}

Internal gravity waves are ubiquitous in the global oceans and are broadly considered to be responsible for a significant proportion of ocean mixing away from boundaries \citep{Garrett1972, Wunsch1975, Hibiya1998}. While one path to the small, viscosity-affected scales that are necessary for mixing arises from the turbulence produced by internal wave breaking \citep{VanHaren2017, MacKinnon2013}, another is the transfer of energy from larger scales to smaller through the route of wave resonances. The triadic resonance instability (TRI) is one such resonance mechanism, in which a single internal wave can become unstable and begin emitting two other internal waves of lower frequency, with all the members of this triad of waves satisfying 
\begin{equation}\label{eqn: resonance equations}
            \wnv_0 = \wnv_1 + \wnv_2, \qquad \omega_0 = \omega_1 + \omega_2, 
\end{equation} where $\wnv$ is the wavenumber vector and $\omega$ is the angular frequency, with subscript $0$ indicating the original or primary wave and $1,2$ indicating the two secondary waves. Early laboratory and theoretical studies established TRI as a pathway for energy transfer and mixing in stratified flows \citep{Davis1967, McEwan1971, McComas1977, McEwan1975}, while later numerical studies demonstrated its role in enhanced oceanic dissipation and mixing \citep{Hibiya2002, MacKinnon2005, Nikurashin2011}. 

For finite-width internal wave beams, there exists an amplitude threshold (for the primary wave forcing) that must be surpassed for TRI to grow \citep{Koudella2006, Bourget2013, Bourget2014, Karimi2014}. \citet{Grayson2022} showed that long-term amplitude fluctuations observed in finite-width-beam experiments were not captured by the existing weakly nonlinear theory. 

The experimental work of \citet{Grayson2024} aimed to reach a triadic state in a finite-width beam beneath  the linear amplitude threshold by injecting a shortlived pulse -- in the form of a vortex ring -- to disturb the primary wavebeam. These experiments demonstrated that perturbing a linearly stable sub-threshold wavebeam with a vortex ring can trigger two distinct sub-threshold triadic states: a long-lasting sustained response (`sustained triadic response', STR), and a short-term response that decays back to the single wave state (`transient triadic response', TTR). These responses suggest that interactions (known to occur in the ocean) between internal wavebeams and turbulent flow structures  may enable wave-wave energy transfer to smaller scales even below the linear TRI threshold, with important consequences for the supply of energy for mixing and the  modelling of oceanic energy budgets. 

In this paper, we investigate further the properties of both TRI and these new sub-threshold responses using sparsity-promoting dynamic mode decomposition (spDMD), which provides an objective criterion for mode selection that is unavailable using standard dynamic mode decomposition (DMD). This reveals additional properties of both the above-threshold and sub-threshold responses. 

This article is composed of the following sections. Section \ref{sec:Experimental procedure} describes the experimental setup and procedures, with \S\,\ref{sec:detection of triads} introducing the spDMD methodology and sensitivity analysis. Results are presented in \S\,\ref{sec:results}, with discussions \S\,\ref{sec:Discussion} and conclusions in \S\,\ref{sec:Conclusions}. 

\section{Experimental procedure}\label{sec:Experimental procedure}
\subsection{Experimental overview}
 Experiments were carried out in the facility used in a number of previous studies \citep{Dobra2019, Dobra2021, Dobra2022, Grayson2021, Grayson2024, Irvine2025}. This facility utilises an 11 m long, 0.48 m deep, and 0.29 m wide Perspex (acrylic) wave tank. The tank was filled with a linear salt stratification to a depth of $0.45 \pm 0.01$ \unit{\metre}, with buoyancy frequency $N = 1.44 \pm 0.06$ \unit{\radian\per\second}, where the uncertainty was measured based on the change in the density profiles that were taken with an aspirating conductivity probe before and after each experiment. The stratification was created with the aid of a pair of computer-controlled gear pumps, using an improved protocol to that used in \citet{Grayson2022}. 

\begin{figure}
    \centering
    \includegraphics[width=0.9\linewidth]{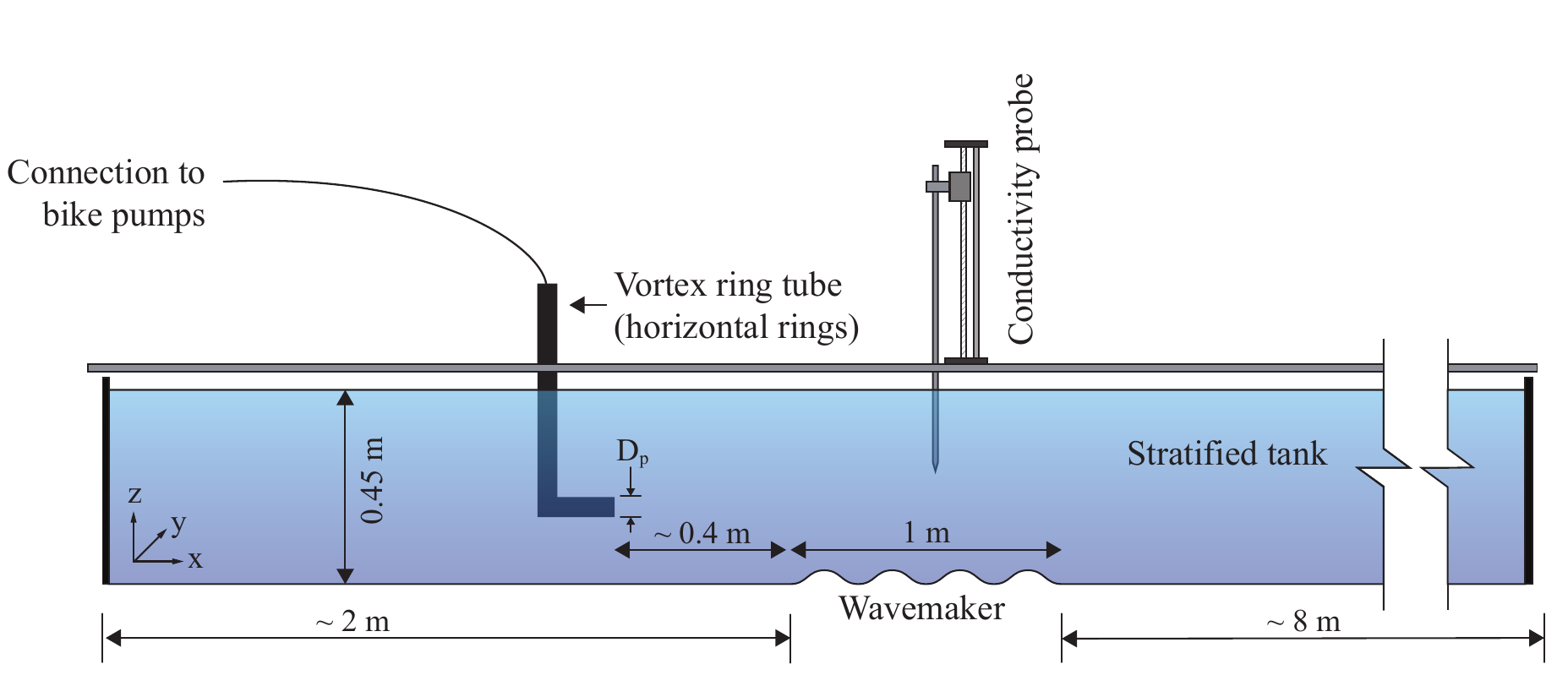}
    \caption{Experimental setup for experiments in the wavetank. The tank is filled with a linear salt stratification and the conductivity probe is used to determine the density profile within the tank. The wavemaker perturbs the bottom of the stratification to produce a quasi-two-dimensional internal wavebeam. Vortex rings are generated by the horizontal L-shaped tube.}
    \label{fig:experimental setup in wavetank}
\end{figure}

Internal waves were generated by perturbing the bottom of the stratification with an arbitrary spectrum wavemaker, known either as the `Magic Carpet' or simply the `wavemaker'. The Magic Carpet is a \SI{1}{\m} long, active, flexible and computer-controlled section of the base of the wavetank (Figure \ref{fig:experimental setup in wavetank}), and can be programmed for forcing frequency $\omega_0$, amplitude $A_0$, and wavelength $\lambda_0$ that vary in both space and time. The high depth-to-width aspect ratio of the tank forms a suitable environment for essentially two-dimensional perturbations from the Magic Carpet (\ie cross-tank variations were negligible). We use the subscript $0$ to indicate the `primary' wave beam induced by displacements of the wavemaker. 

Full details of the construction, computer hardware and software of the Magic Carpet are available in \citet{Dobra2018}, while details of previous scientific usage can be found in \citet{Dobra2019, Dobra2022}, \citet{Grayson2021} and \citet{Grayson2022, Grayson2024}. Further experimental details are in available in \citet{Irvine2025} \S\,4.2 and 4.3.

\subsection{Internal wave forcing}
The Magic Carpet was programmed to generate a vertical displacement of 
\begin{equation}
    \label{eqn: wave beam envelope}
    z = h(x,t) = \begin{cases}
    \Re\big(f(t) \hspace{0.6mm} e^{i \pwx x} \cos^2\big(\frac{x-B}{8\pi^2}\big)\big), &  \qquad  A < x < B,  \\
    \Re\big(f(t) \hspace{0.6mm} e^{i \pwx  x} \big),  & \qquad   B < x < C,   \\
    \Re\big(f(t) \hspace{0.6mm} e^{i \pwx x} \cos^2\big(\frac{x-C}{8\pi^2}\big)\big), &   \qquad C < x < D, \\
    0, & \qquad   \textmd{elsewhere}, \\
    \end{cases}
\end{equation} where $\Re(\cdot)$ gives the real component of the expression. and $\pwx$ is the horizontal component of the primary wave vector $\pwv = (\pwx, \pwz) =($0.05, -0.06$)$ mm$^{-1}$. The origin $x = 0$ was defined as the leftmost side of the wavemaker when viewed from the perspective of the camera (as shown in Figure \ref{fig:experimental setup in wavetank}). The locations $A, B, C, D$ were (respectively) $9\pi/\pwx$, $11\pi/\pwx$, $15\pi/\pwx$, $17\pi/\pwx$. This gives a full horizontal wavelength for the primary beam of $\lambda_{x_0}$ = 2$\pi/\pwx$ $=$ 125 \unit{\mm}. 

The time-varying amplitude of the forcing $f(t)$ is then described as
\begin{equation}
\label{eqn: constant amp forcing}
f(t) = \begin{cases}
0, 	& \hspace{4mm} t \leq 0 \: \mathrm{s}, \\ 
\aswamamp \hspace{0.6mm}\big(\frac{t}{30}\big)e^{-i\omega_0 t},  & \hspace{4mm} 0 \leq t \leq 30 \: \mathrm{s}, \\
\aswamamp \hspace{0.6mm}e^{-i\omega_0 t}, & \hspace{4mm} 30 \: \mathrm{s} \leq t \leq t_\textrm{end} -30 \: \mathrm{s} \:, \\
\aswamamp \hspace{0.6mm}\big(\frac{t_\textrm{end} -t}{30}\big)e^{-i\omega_0 t}, & \hspace{4mm} t_\textrm{end} -30 \: \mathrm{s} \leq t \leq t_\textrm{end} \:, 
\end{cases} \quad
\end{equation}
where $\omega_0$, $\aswamamp$ and $t_\textrm{end}$ are (respectively) the forcing frequency (0.95 rad s$^{-1}$), the nominal forcing amplitude of the primary beam (3.0 to 3.9 mm), and the end time of the experiment (3660 seconds). We will refer to the mode produced by the primary wave as $\wave_0$, and additional modes arising from TRI or wave--wave interaction as $\wave_j$ with $j = 1,2,3,4,5,$ and so forth. This incorporates both the frequency as well as the spatial structure of the spDMD mode, and allows reference to the mode without specifying the form of the disturbance (wave, evanescent wave, oscillatory signal, \etc). For this and following sections, we have non-dimensionalised all frequencies by the buoyancy frequency $N$, using $\omega_j$ for the dimensional frequency and $\Omega_j = \omega_j/N$ as the non-dimensional frequency. We also non-dimensionalise time by the period $T_0$ of the primary wave $\wave_0$ to give $\tau = t/T_0$. 

\subsection{Linear threshold determination}
While a theoretical estimate of the linear threshold for the onset of TRI can be made, we chose to account for any non-ideal properties of our experiments by making direct measurements. The linear threshold was determined from experiments containing solely the forced primary wavebeam $\wave_0$; these are defined more fully in \S\,\ref{sec:detection of triads} with outcomes in \S\,\ref{sec:comparison of results}. If secondary waves were observed to develop, the $\wave_0$ amplitude was designated as being above-threshold. We determined the range of amplitudes over which the transition occurred from absence to presence of TRI by plotting the response type against amplitude. 
If TRI were to be triggered, the secondary waves would only be seen once the instability had grown above the diagnostic noise baseline. In our observations this manifests as a delay between the actuation time of the primary wave and the time after which the secondary waves are detectable. 

\subsection{Vortex ring generation}\label{subsec:vortex ring generation}

Vortex rings were used in some experiments as the means to perturb the primary internal wave beam. Vortex rings were chosen to be a  controllable and relatively reproducible analogue for turbulence and have the useful property that they would not in isolation produce significant mixing, and thus preserve the usable life of the background stratification. One of the important dimensionless numbers for vortex rings is their formation number, defined as 
\begin{equation}\label{eqn:chap3 formation number}
    F_p = L_p/D_p, 
\end{equation} where $L_p$ is the length of the piston stroke that produces the ring and $D_p$ is the diameter of the vortex-ring tube. For values of $F_p \lesssim 4.0$, an isolated spheroidal ring is produced, whereas for values of $F_p > 4.0$ a trailing jet forms behind the ring \citep{Gharib1998}. We used $3.6 \leq F_p \leq 4.0$ for all experiments to eliminate the complications of a trailing jet, aiming to deliver a discrete (albeit propagating) pulse of vorticity into the wavebeam. 

To describe the ring evolution, we introduce two additional dimensionless parameters, the ring transition point $\mathring{x}_t$ and the ring endpoint $\mathring{x}_e$. The ring transition point is related to the ring Froude number, given by
\begin{equation}
    Fr = \frac{U}{Na},
\end{equation} where $U$ is the instantaneous horizontal velocity of the ring, $N$ is the buoyancy frequency of the stratification, and $a$ is the characteristic length scale, taken here to be the 20 \unit{\mm} radius of the vortex ring tube outlet (\ie $a = D_p/2$). \citet{Scase2006b} noted that as a ring travels in a stratified fluid and momentum diffuses outwards, the Froude number decays until $Fr \thicksim 1$ whereupon the ring no longer possesses sufficient kinetic energy to exchange for the potential energy required to lift fluid up and around the ring. Here, we refer to this point as $x_t$ and we nondimensionalise this by 
\begin{equation}
    \mathring{x}_t = {x_t}/\lambda_{x_0},
\end{equation} where $x_t$ is then horizontal distance (measured from the tube opening) at the $Fr=1$ transition and $\lambda_{x_0}$ is the wavelength of a primary internal wavebeam towards which a vortex ring is directed. The non-dimensional endpoint of the vortex ring path, $\mathring{x}_e = x_e/\lambda_{x_0}$, is defined as the location where the structure is so incoherent as to be indistinguishable from diagnostic noise. The corresponding times in the experiments for these occurrences are given as $\tau_t$ and $\tau_e$. The initial velocities of the rings were varied between 36 \unit{\mm\per\second} and 106 \unit{\mm\per\second}, generating $1.28 < Fr < 3.25$ and $1450 < Re < 4240$. Details of the image processing and calculations for ring velocity and height are given in the appendix of \citet{Grayson2024}. 

\subsection{Wave visualisation}\label{subsec:wave visualisation}
The internal waves in these experiments were visualised using synthetic schlieren, a non-intrusive method first proposed by \citet{Dalziel1998} and further developed in \citet{Sutherland1999}, \citet{Dalziel2000}, and \citet{Dalziel2007}. The method relates density perturbations to refractive index changes due to movement of isopycnals within the tank, allowing calculation of the amplitudes of the internal wave. 

From the processed synthetic schlieren images, we compute the line-of-sight mean of the gradient vector of the density perturbation field $\rho$, producing 
\begin{equation}
\boldsymbol{\beta} = (\beta_x, \beta_z) =  \frac{g}{N^2\varrho_0}\bigg(\frac{\partial{\rho}}{\partial x}, \frac{\partial{\rho}}{\partial z}\bigg),
\label{eqn: output from ss}
\end{equation} where $\boldsymbol{\beta}$ is non-dimensionalised for convenience, using gravitational acceleration $g$, the buoyancy frequency $N$, and a reference density $\varrho_0$.  

The amplitude of each beam in a triad then is given by
\begin{equation}
    \xi_p(x,z) = \int_x\beta_x\mathrm{d}x + \left.\int_z\beta_z\mathrm{d}z\right|_{x=0}+C,
\end{equation}
where $C$ is set to zero if the perturbation field has vanishing mean. We assume an oscillatory form of the vertical component of the fluid displacement $\xi_p = \tilde{\xi}_pe^{\textrm{i}(\boldsymbol{k}_p \cdot \boldsymbol{x} + \omega_p t)}$, where $\tilde{\xi}_p$ is a slowly varying complex amplitude field, distinct from the fast oscillations of the primary wave described by the exponential. Initially $\tilde{\xi}_p$ has not been spatially filtered, so the domain contains multiple wavevectors with a common wavenumber that differ in their direction due, for example, to reflections from the free surface. To filter out unwanted reflections we employ the Hilbert Transform \citep{Mercier2008} and thus isolate the quadrant in Fourier space containing the beam of interest. After taking the root mean square of the complex output from the Hilbert transform, we spatially average over the whole field of view, producing the filtered quantity $\langle|\breve{\xi}_p|\rangle_w$. For full details of processing, see \citet{Grayson2024}. 

Our experiments used a baseline frame rate of 1 frame per second (fps), which is more than sufficient to resolve the period $T_0= 2\pi / \omega_0 \approx$ 6.62 $\si{\second}$ of the primary wavebeam $\wave_0$ and the higher frequencies these experimental reveal. For experiments containing a vortex ring perturbation in addition to $\wave_0$, we introduced a 20 fps frame rate during the lifespan of the vortex ring then returned to 1 fps for the remainder of the experiment, limiting data storage requirements to 50--60 GB per (hour-long) experiment. This higher frame rate was necessary to capture the rapid ring evolution of the ring in order to determine the Froude number $Fr$, and transition and end points ($\mathring{x}_t$ and $\mathring{x}_t$, respectively) defined below.

\section{Detection of triads}\label{sec:detection of triads}

Using spDMD, we uncover higher-frequency oscillatory signals that were not identified in the earlier analysis of \citet{Grayson2024}. We analyse two subsets of the experimental dataset of \citeauthor{Grayson2024}: Set \setw consists of 35 wave-only experiments used to establish the linear TRI threshold, and Set \setb consists of 29 experiments containing an interaction between a forced internal wavebeam and a horizontally-fired vortex ring perturbation. The synthetic schlieren image sequences corresponding to $x$ and $z$ gradients of the density perturbation field were preprocessed by cropping extraneous areas in each video-frame then stacked with the $z$-gradient image abutting that for the $x$-gradient (see \citet{Grayson2022}), then, treating each combined image as a matrix, a hard threshold is applied to the singular value spectrum of this matrix to eliminate pixel-scale noise that plays no dynamical role. The sequence of combined images was then analysed using spDMD. 

\subsection{Applying sparsity-promoting dynamic mode decomposition}\label{subsec: applying sparsity-promoting dmd}

Previous analyses of these experiments in \citet{Grayson2024} used the standard approach to DMD in which the number of retained modes is determined by a user-defined rank truncation. In noisy experimental data, however, fixed-rank truncation can suppress weak but coherent structures or retain modes dominated by large-amplitude, short-duration noise. We therefore revisit the data using spDMD \citep{Jovanovic2014} to determine whether additional dynamically significant features were excluded by fixed-rank truncation. 

This technique selects a sparse subset of DMD modes by seeking to minimise a reconstruction error that we define to be the difference between a sparse and a full-rank superposition of DMD modes. Rather than prescribing the number of modes beforehand, spDMD returns an optimal set of modes by balancing with a Lagrange multiplier the reconstruction accuracy against sparsity. 

Here, spDMD is applied to each \SI{20}{\second} analysis window of the preprocessed synthetic schlieren images, and the resulting modal frequencies, amplitudes, and spatial structures are analysed. Appendix \ref{appendix:spdmd matrices} summarises the notation used later in the paper when referring to spDMD outputs; the full mathematical details are given by \citet{Jovanovic2014} and reviewed by \citet{Schmid2022}.  
 
\subsection{Selecting reconstruction error limit}

\begin{figure}
    \centering
    \includegraphics[width=0.8\linewidth]{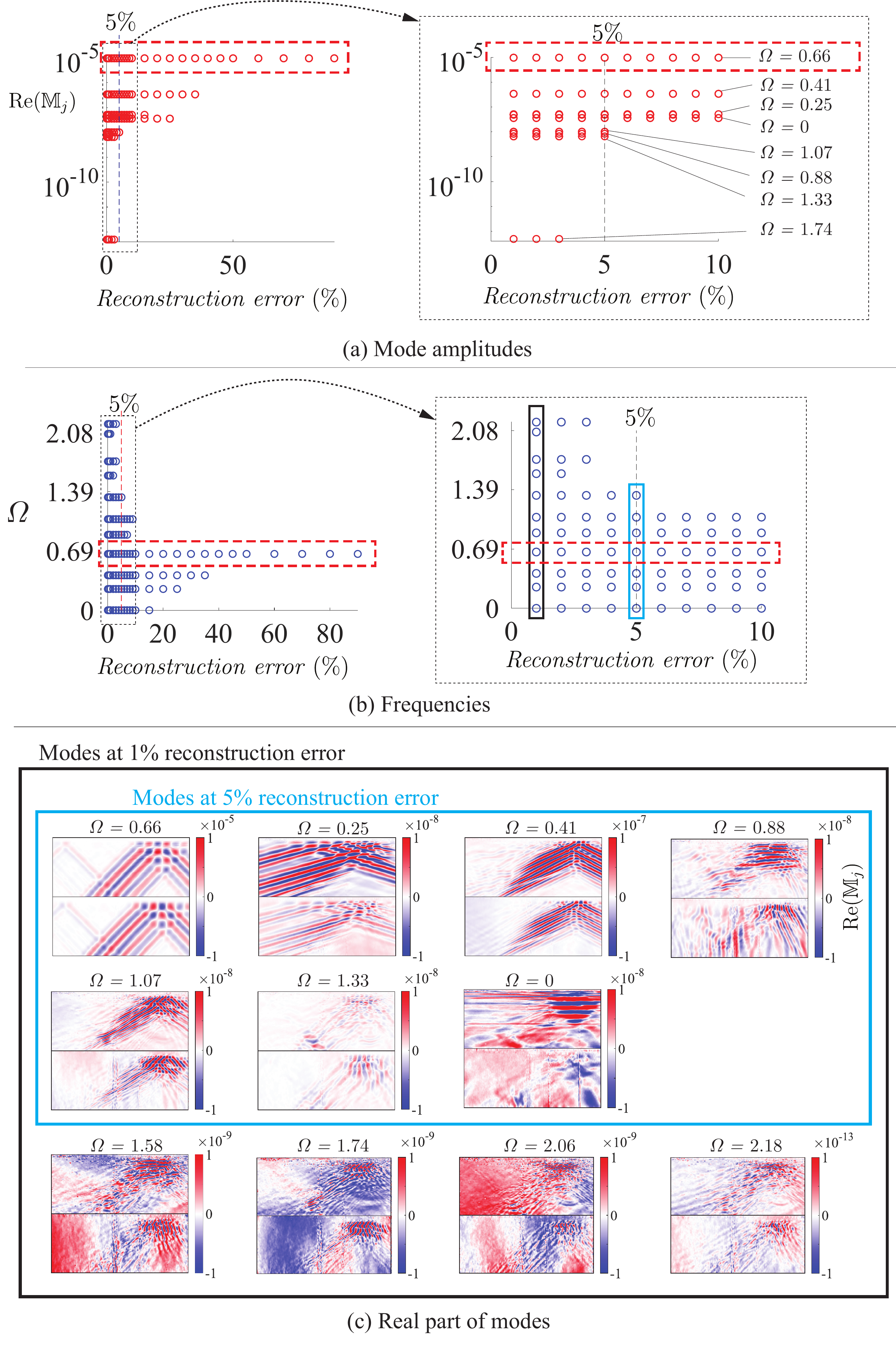}
    \caption{
    Analysis of mode selection by spDMD algorithm on case of TRI, for reconstruction-error tolerances between 1\% and 90\%. In (a) and (b) the right plot is an enlarged reproduction of the grey box on the left plot. Dashed red boxes correspond to the dominant frequency $\Omega_0$ of the primary beam. For each sampled reconstruction-error tolerance, the number of DMD modes detected is found by counting circles aligned vertically, In (c) the spatial distribution of signal is given. Modes inside the blue box are those found with a 5\% tolerance; those only detectable with a 1\% tolerance are within the black box but outside the blue box.}
    \label{fig:TRI sensitivity check}
\end{figure}

Using spDMD requires a user-defined choice of tolerance for the reconstruction error. We aim to set a tolerance that captures the dominant features of the flow while excluding noisy and dynamically unimportant artefacts. To determine an appropriate tolerance, we carried out sensitivity analyses over a broad range of values on data that were representative of our experimental archive. For cases featuring TRI, a stable set of dynamically significant modes emerged for tolerances between $4\%$ and $10\%$ (Figure \ref{fig:TRI sensitivity check}). Primary and secondary TRI modes were consistently retained, alongside several higher-frequency components. Within this range, varying the tolerance changed the number of retained modes but did not appreciably alter the frequencies, amplitudes, or spatial structure of the modes; in particular, the mode amplitudes were unchanged to three decimal places. 

Below about $4\%$, additional modes appeared (those outside the blue box in Figure \ref{fig:TRI sensitivity check}) but these were typically an order of magnitude smaller than the non-$\wave_0$ modes and lacked coherent spatial structure, suggestive of having reached a noise threshold. We therefore selected a 5\% reconstruction error for our analysis, as an acceptable balance between reconstruction accuracy and sparsity. A sensitivity analysis for a representative transient case is presented in Appendix \ref{Appendix:sensitivity on TTR}, which supports our choice. In selecting modes by reconstruction-error tolerance rather than by singular value truncation, spDMD allows the number of retained modes to vary with the complexity of the flow itdelf, and provides an objective measure of error. 

% ***** TRI sensitivity example *****
% ***********************************

\FloatBarrier
%%%%%%%%%%%%%%%%%%%%%%%%%%%%%%%%%%%%%%%%%%%%%%%%%%%%%%%%%%%

\section{Results}\label{sec:results}

We begin with representative case studies of the three observed response types (\S\,\ref{subsec:characterisation of regimes}), then compare the newly identified higher-frequency features across these cases (\S\,\ref{subsec:non-propagating disturbances}), and then fully summarise the results over the full dataset in \S\,\ref{sec:comparison of results}.

\subsection{Case studies}\label{subsec:characterisation of regimes}

We present case studies from experimental Sets \setw and \setb, illustrating the three observed triadic responses --- TRI, STR, and TTR. For each case, we examine the evolution of frequency and amplitude evolution, together with representative mode structures. The amplitude boundaries for the response types are shown in Table \ref{tab:table of wave amplitudes for each regime} and are derived from the full set of experiments shown in \S\,\ref{sec:comparison of results}. 

\begin{table}[h]
    \centering
    \begin{tabularx}{0.6\linewidth}{c >{\raggedleft\arraybackslash}X}
        \toprule
        Response & Amplitude range\\
        \midrule
         TRI & $\pwx\langle|\breve{\xi}_0|\rangle_w \gtrsim 0.0218$ \\
         STR & $0.0180 \lesssim \pwx\langle|\breve{\xi}_0|\rangle_w \lesssim 0.0218$ \\
         TTR & $0.0164 \lesssim \pwx\langle|\breve{\xi}_0|\rangle_w  \lesssim 0.0180$ \\
         No response & $\pwx\langle|\breve{\xi}_0|\rangle_w \lesssim 0.0179$ \\
         \bottomrule
    \end{tabularx}
    \caption{Primary wave input amplitude ranges for the triadic responses. The linear threshold is at $\pwx\langle|\breve{\xi}_0|\rangle_w \approx 0.0218$}
    \label{tab:table of wave amplitudes for each regime}
\end{table}

\subsubsection{Triadic resonance instability}\label{subsec:TRI}

We first consider an above-threshold TRI case, shown in Figure \ref{fig:wiggle plots, a}, for which the primary wave has a dimensionless amplitude of $\pwx\langle|\breve{\xi}_0|\rangle_w \approx 0.0223$ and no vortex ring perturbation is present. The figure shows the evolution of the frequencies recovered by spDMD, with marker colour indicating dimensionless mode amplitude. Two secondary waves ($\wave_1$ and $\wave_2$) become visible about 300 seconds into the experiment $(\tau \approx 45)$, and persist until the forcing of the primary wave ceases. These secondary waves fluctuate slightly in frequency and amplitude over this duration, but their frequency fluctuations remain paired such that they always satisfy the temporal relation for resonance \eqref{eqn: resonance equations} with the forced primary wave $\wave_0$. Over longer times, these frequency fluctuations appear quasi-periodic, with standard deviations of 0.007 for both, and period $\tau \approx 140$ to $180$, giving $T_\text{fluc} \approx 265$ s. 

In addition to these three waves, Figure \ref{fig:wiggle plots, a} shows additional signal at dimensionless frequencies of $\Omega \approx 0.89$ and $\Omega \approx 1.07$, appearing approximately 460 to 480 seconds ($\tau \approx 69$ to $72$) into the experiment. We label these as $\wave_3$ and $\wave_4$, respectively, noting that their frequencies are close to those that would arise from
\begin{equation*}
        \Omega_3  = \Omega_0 + \Omega_1 \approx 0.66  + 0.24 \approx 0.90, 
\end{equation*} and
\begin{equation*}
        \Omega_4 = \Omega_0 + \Omega_2 \approx 0.66  + 0.41 \approx 1.07. 
\end{equation*} Also present is signal denoted by $\wave_5$ with $\Omega_5 = 2\Omega_0$. This, interestingly, seems to appear from the beginning of $\wave_0$ at 200 seconds ($\tau \approx 30$) and remains nearly constant. A steady signal of $\Omega=0$, which we designate $\wave_6$, is also present but is not shown in Figure \ref{fig:wiggle plots, a}. We examine these additional frequency components in greater detail in \S\,\ref{subsec:non-propagating disturbances}.  

% ***** TRI wiggle plot *****
\begin{figure}
    \centering
    \includegraphics[width=12cm]{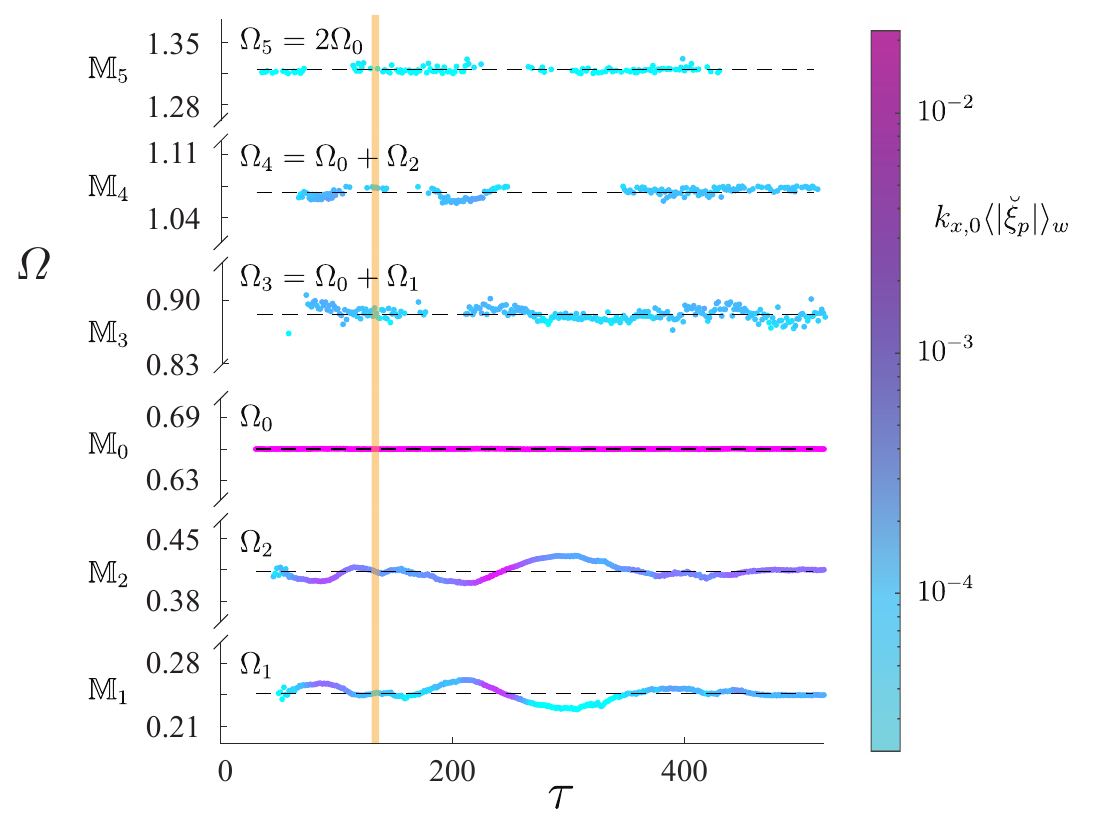}
    \caption{The evolution of frequency over time in an experiment containing TRI ($\pwx\langle|\breve{\xi}_0|\rangle_w \approx 0.0223$). The non-dimensional amplitude $\pwx\langle|\breve{\xi}_p|\rangle_w$ of each wave $p$ shown by the colour of the markers. The horizontal axis gives non-dimensional time $\tau = t/T_0$. An initial period of $\tau < 30$ is excluded, covering the  $30 T_0 \approx 200$ \unit{\second} of the Magic Carpet's ramp up. Note the vertical axis is discontinuous to focus on frequency regions in which consistent signal is found, and dashed grey lines denote the mean frequency value of the signal in each individual band. These mean frequencies are $\Omega_0 = 0.66$ for the forced primary wave, and $\Omega_1 = 0.25$, $\Omega_2 = 0.41$, $\Omega_3 = 0.88$, $\Omega_4 = 1.07$, and $\Omega_6 = 0$. The cream-coloured highlighting marks the $\tau = 169$ DMD window from which the TRI modes are calculated.}
    \label{fig:wiggle plots, a}
\end{figure}

\begin{figure}
    \centering
    \includegraphics[width=1\linewidth]{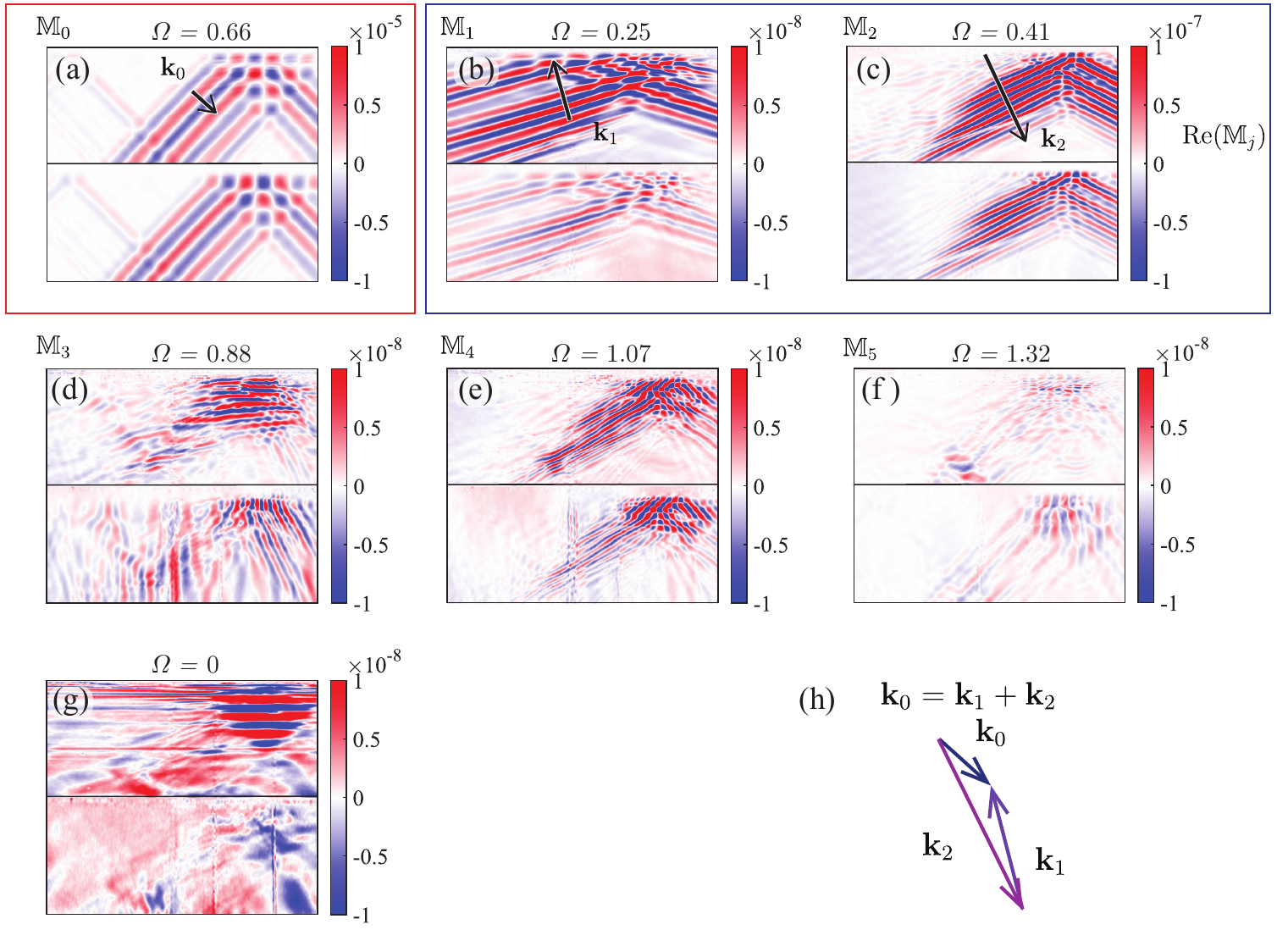}
    \caption{Modes corresponding to the frequencies from above-threshold $(\wnxa{0}\langle |\breve{\xi}_0|\rangle_w\approx 0.0223)$ TRI case shown in Figure \ref{fig:wiggle plots, a}. The frequencies are (a) $\wave_0:\Omega = 0.66$, (b) $\wave_1:\Omega = 0.25$, (c) $\wave_2:\Omega = 0.41$, (d) $\wave_3:\Omega = 0.88$, (e) $\wave_4:\Omega = 1.07$, (f) $\wave_5:\Omega = 1.33$ and (g) $\wave_6:\Omega = 0$. The modes are from the interval centred on $\tau = 169$, marked by the cream highlighting in Figure \ref{fig:wiggle plots, a}, with the colour bar for each image indicating the amplitude of the real part of the spDMD mode. Note that the scales for the colour bars differ between images. The primary wave is boxed in red, with the two secondary waves that form the $\wnv_0 = \wnv_1 + \wnv_2$ triad boxed in blue. Wavevectors drawn on (a) to (c) are calculated and scaled from the wavelengths observed in each image. The $\wnv_0 = \wnv_1 + \wnv_2$ triad is shown graphically in (h).}
    \label{fig:TRI modes}
\end{figure}

Figure \ref{fig:TRI modes} shows the real part of the mode structure corresponding to the observed frequencies. Black arrows indicate the wavenumber vectors $\wnv_j$, calculated from the observed mode wavelengths and oriented perpendicular to the lines of constant phase. The primary wave $\wave_0$ is shown in Figure \ref{fig:TRI modes}a (red box), while the two secondary waves $\wave_1$ and $\wave_2$ are shown in Figure \ref{fig:TRI modes}(b) and (c)  (blue box). Combining $\wnv_0$, $\wnv_1$, and $\wnv_2$ graphically produces the wavenumber vector diagram in Figure \ref{fig:TRI modes}(h), showing $\wnv_0 = \wnv_1 + \wnv_2$. By comparing the wavenumbers and the angles of the modes with the dispersion relation, it becomes clear that these three signals satisfy both the dispersion and resonance relations and are therefore a triad of propagating waves. 

% ***** TRI modes *****

\begin{figure}
    \centering
    \includegraphics[width=12cm]{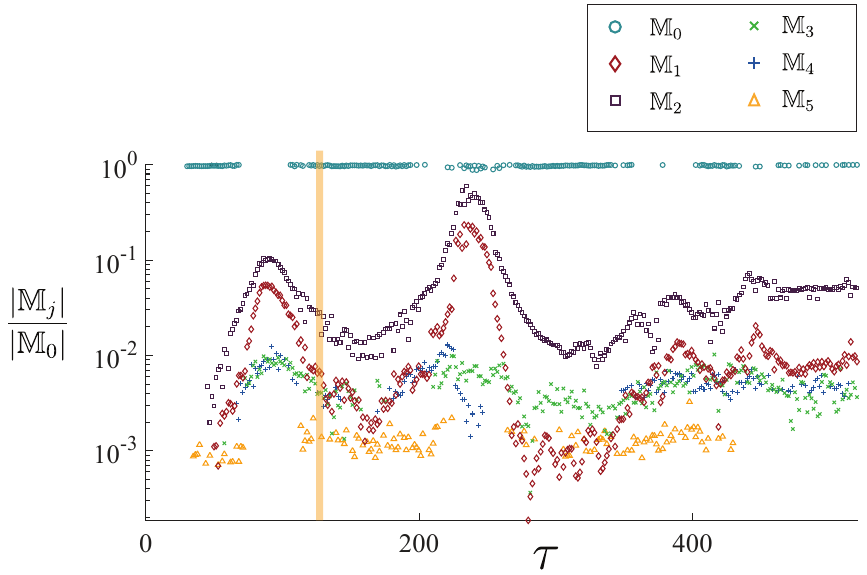}
    \caption{Normalised amplitudes of the waves/disturbances identified from the TRI case in Figure \ref{fig:wiggle plots, a}. All waves are normalised by the modulus of the complex amplitude of the primary wave for each window. The cream-coloured vertical strip at $\tau = 169$ indicates the DMD window from which the TRI mode examples are calculated.} 
    \label{fig:TRI amplitude plot}
\end{figure}

Figure \ref{fig:TRI amplitude plot} shows the evolution of the modulus of the complex mode amplitudes, calculated for each 20-second DMD window and normalised by the modulus of the complex mode amplitude of the primary wave. For the majority of the experiment, the $\wave_1$ and $\wave_2$ waves have the largest mode amplitudes of the non-primary waves, though both exhibit fluctuations. In particular, the amplitude of the $\wave_1$ wave briefly falls below that of $\wave_3$, $\wave_4$, and $\wave_5$ around $\tau \approx$ 300 before recovering. Both the $\wave_3$ and $\wave_4$ components also fluctuate in a quasi-periodic manner, approximately in-phase with their corresponding parent secondary waves, \ie $\wave_3$ with $\wave_1$ and $\wave_4$ with $\wave_2$. 

% ***** TRI amplitude plot ********

Our analysis therefore indicates that TRI contains not only the propagating wave triad formed of $\wave_0$, $\wave_1$, and $\wave_2$, but also the higher frequency oscillatory signals of $\wave_{3, 4, 5}$ and the zero-frequency component $\wave_6$. While the secondary waves of $\wave_1$ and $\wave_2$ were also found by \citet{Grayson2024} in their analysis of TRI, the additional higher frequency `tertiary' combinations $\wave_{3,4,5}$ were not detected because the fixed-rank truncation in their DMD analysis eliminated them. 

%%%%%%%%%%%%%%%%%%%%%%%%%%%%

\subsubsection{Sustained triadic response}\label{subsec:STR}
We next consider sustained triadic response (STR), shown in Figure \ref{fig:wiggle plots, b}. In this below-threshold case, the primary wave has dimensionless amplitude $\pwx\langle|\breve{\xi}_0|\rangle_w \approx 0.0195$. No secondary waves were detected before vortex-ring perturbation, which is introduced at
$\tau \approx 48$. Unlike the TRI cases, the vortex ring was visible as broad-spectrum frequency scatter around $\tau = 48$ to $52$. Following this perturbation, secondary waves $\wave_1$ and $\wave_2$ appear rapidly, converge towards stable frequencies, and then persist until the end of the experiment while continuing to satisfy the temporal resonance relation with the primary wave $\wave_0$. Quasi-periodic frequency fluctuations were also evident, with similar magnitude and period to those in the TRI. 

The $\wave_3$ signal emerged on a similar time scale to $\wave_1$ and $\wave_2$, showing short-period fluctuations ($\tau \approx 23$) until $\tau \approx 151$. In contrast, $\wave_4$ emerged after $\tau\approx 83$, when $\wave_1$ and $\wave_2$ were relatively stable in frequency. As observed in the TRI case, the emergence of $\wave_5$ is coincident with $\wave_0$, but is only intermittently detected. This intermittency is common for the higher frequency signals $\wave_{3,4,5}$ in most STR cases. 

A plausible explanation for this weaker detection in below-threshold cases is simply that amplitudes of already-weak modes are small relative to the noise baseline when compared with above-threshold excitation. Since higher-frequency signals draw power from the primary and secondary waves, a direct consequence of a smaller amplitude in $\wave_0$ is less power available in $\wave_{0,1,2}$ to power further interactions. For reference, the TRI example had $\wnxa{0}\langle |\breve{\xi}_0|\rangle_w\approx 0.0223$, compared to $\pwx\langle|\breve{\xi}_0|\rangle_w \approx 0.0195$ for this STR example. 

\begin{figure}
    \centering
    \includegraphics[width=12cm]{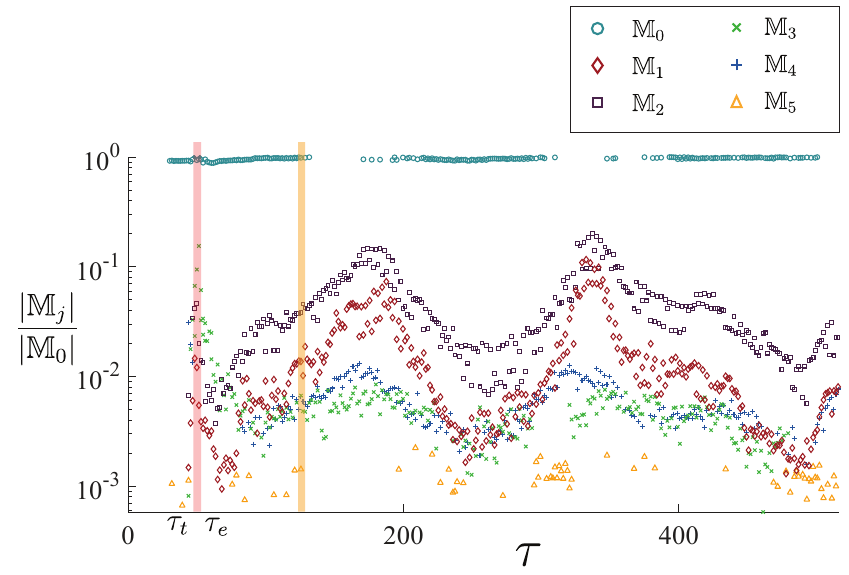}
    \caption{Normalised amplitudes of the waves/disturbances identified from the STR case in Figure \ref{fig:wiggle plots, b}. All waves are normalised by the real part of mode amplitude of the primary wave for each window. The cream-coloured highlighting indicates the $\tau = 169$ DMD window from which the TRI mode examples are calculated, while the pink highlighting indicates $\tau_t$ and $\tau_e$ for the vortex ring.}
    \label{fig:STR amplitude plot - vortex rings}
\end{figure}

Figure \ref{fig:STR amplitude plot - vortex rings} illustrates that the amplitude fluctuations for $\wave_1$ and $\wave_2$ are approximately out of phase, and that the secondary and higher-frequency signals otherwise behave similarly to those in the TRI case. The mean values of STR $\Omega_1$ and $\Omega_2$ differ from those in TRI only by $\pm 0.01$, while continuing to satisfy the temporal resonance relation with $\Omega_0$. The corresponding STR mode structures are almost identical to the TRI modes in Figure \ref{fig:TRI modes}, and are shown in Appendix \ref{Appendix: STR and TTR modes}. 

% ***** STR wiggle plot *****
\begin{figure}
    \centering
    \includegraphics[width=12cm]{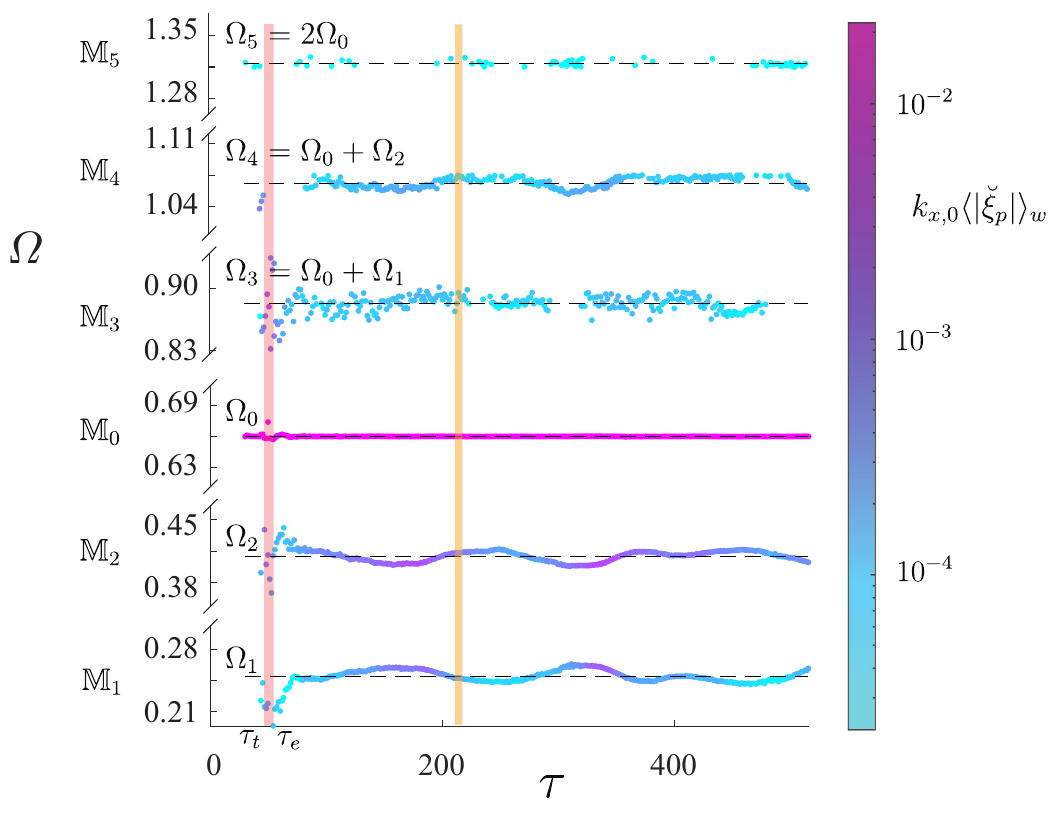}
    \caption{Evolution of frequency over time of an experiment containing STR ($\pwx\langle|\breve{\xi}_0|\rangle_w \approx 0.0195$). The non-dimensional amplitude $\pwx\langle|\breve{\xi}_p|\rangle_w$ of each wave $p$ shown by the colour of the circles. Note the vertical axis is discontinuous to focus on frequency regions where signal is found, and dashed grey lines denote the mean frequency value of the signal in each individual band. These mean frequencies are $\Omega_0 = 0.66$ for the forced primary wave, and $\Omega_1 = 0.26$, $\Omega_2 = 0.40$, $\Omega_3 = 0.88$, $\Omega_4 = 1.08$, and $\Omega_6 = 0$. The cream-coloured vertical strip marks the $\tau = 211$ DMD window from which the modes in Appendix \ref{Appendix: STR and TTR modes} are calculated, whereas the pink vertical strip indicates the duration between $\tau_t$ and $\tau_e$ for the vortex ring. The colour scale is identical to that of Figure \ref{fig:TRI modes}.}
    \label{fig:wiggle plots, b}
\end{figure}

% % ****** STR amplitude plot *******

\subsubsection{Transient triadic response}\label{subsec:TTR}
We finally consider a transient triadic response that decays back to a single-wave state before the end of the experiment, shown in Figure \ref{fig:wiggle plots, c} with a primary wave amplitude of $\pwx\langle|\breve{\xi}_0|\rangle_w \approx 0.0170$. As applied in the STR case, a vortex ring perturbated the primary beam at $\tau \approx 48$, after which the secondary waves $\wave_1$ and $\wave_2$ rapidly became visible. Unlike STR, however, these secondary waves persist only for a limited time: the $\wave_2$ wave was no longer detected by $\tau\approx 165$, followed by $\wave_1$ at $\tau \approx 188$. 

When present, $\wave_1$ and $\wave_2$ exhibit matched frequency fluctuations that combine to maintain the temporal resonance relation in \eqref{eqn: resonance equations}. The duration of the detected secondary-wave signal is approximately half of the long-period quasi-periodic fluctuation seen in TRI and STR, consistent with observing only part of a cycle. We therefore hypothesise that a longer-lived TTR might show a similar fluctuation period and frequency range to the TRI and STR cases.

Among the higher-frequency components, only a very brief appearance of $\wave_3$ is observed, localised to the time directly after the vortex-ring perturbation. This, however, disappeared before $\wave_1$ and $\wave_2$ converge towards their respective relatively stable frequencies. The $\wave_{4,5,6}$ modes were either absent or similarly short-lived.

\begin{figure}
    \centering
    \includegraphics[width=0.75\linewidth]{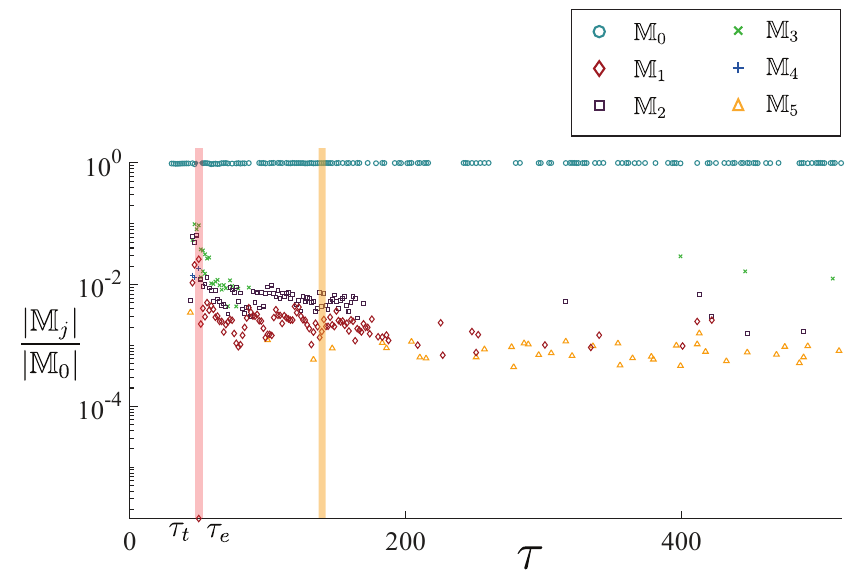}
    \caption{Normalised amplitudes of the waves/disturbances identified from the TTR cases in Figure \ref{fig:wiggle plots, c}. All waves are normalised by the modulus of the complex amplitude of the primary wave for each window. The cream vertical strip marks the $\tau = 167$ DMD window from which the mode shapes (in Appendix \ref{Appendix: STR and TTR modes}) are calculated, while the pink vertical strip indicates $\tau_t$ and $\tau_e$ for the vortex ring.}
    \label{fig:TTR amplitude plot - vortex ring}
\end{figure}

Figure \ref{fig:TTR amplitude plot - vortex ring} confirmed the transient nature of the response and shows fluctuations, in $\wave_1$ particularly, that were also evident in the STR and TRI cases. The corresponding TTR modes are therefore limited to $\wave_0$, $\wave_1$, and $\wave_2$, and are nearly identical to the corresponding modes in TRI and STR. These are shown in Appendix \ref{Appendix: STR and TTR modes}. 

We note that using a more stringent reconstruction-error tolerance (\eg $1\%$) does lead to the detection of $\wave_{3,4,5,6}$ in TTR cases. However, the corresponding normalised moduli of the complex mode amplitudes $(\frac{|\wave_j|}{|\wave_0|})$ were smaller than $\mathcal{O}(10^{-3})$ and close to the diagnostic noise baseline. This lower amplitude, in combination with the short duration of the response, may explain why additional $\wave_{3,4,5,6}$ components in TTR are not retained at this reconstruction-error tolerance. The power available in the secondary waves is already lower than in TRI and STR, and there is less time for that power to be drawn into amplifying the higher-frequency components.

% ***** TTR wiggle plot *****
\begin{figure}
    \centering
    \includegraphics[width=12cm]{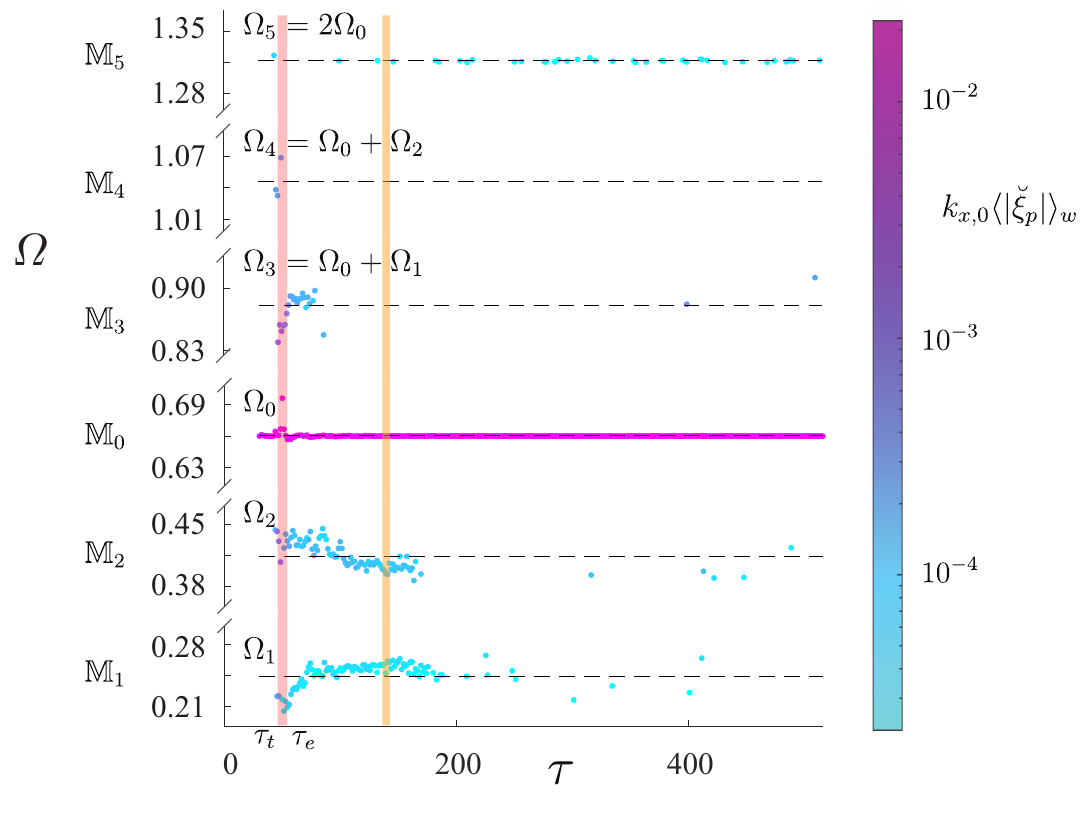}
    \caption{Evolution of frequency over time in an experiment containing vortex ring TTR ($\pwx\langle|\breve{\xi}_0|\rangle_w \approx 0.0170$). The non-dimensional amplitude $\pwx\langle|\breve{\xi}_p|\rangle_w$ of each wave $p$ shown by the colour of the circles, which are on the same scale as Figure \ref{fig:wiggle plots, a}. Note the vertical axis is discontinuous to focus on frequency regions with signal, and dashed grey lines denote the mean frequency value of the signal in each individual band. These mean frequencies are $\Omega_0 = 0.66$ for the forced primary wave, and $\Omega_1 = 0.25$ and $\Omega_2 = 0.41$. The cream vertical strip marks the $\tau = 167$ DMD window from which the modes are calculated (Appendix \ref{Appendix: STR and TTR modes}), while the pink vertical strip indicates $\tau_t$ and $\tau_e$ for the vortex ring.} 
    \label{fig:wiggle plots, c}
\end{figure}

%%%%%%%%%%%%%%%%%%%%%%%%%%%%%%%%%%%%%%%
\FloatBarrier
\subsection{Non-propagating wave--wave interactions}\label{subsec:non-propagating disturbances}

An important and novel finding from our new analysis, above, is the presence of the higher-frequency signals $\wave_{3, 4, 5}$  in TRI and STR cases, beyond the propagating triad of $\wave_{0,1,2}$. We now examine these signals in more detail, using the TRI case as a representative example. 
%We have selected spDMD windows during which $\frac{\partial \Omega_{1,2}}{\partial t}\approx 0$ 
Our aim is to determine whether $\wave_{3,4,5}$ can be interpreted as propagating internal waves, or whether they are better understood as confined, non-propagating products of further wave--wave interactions. 

\begin{figure}
    \centering
    \includegraphics[width=0.6\linewidth]{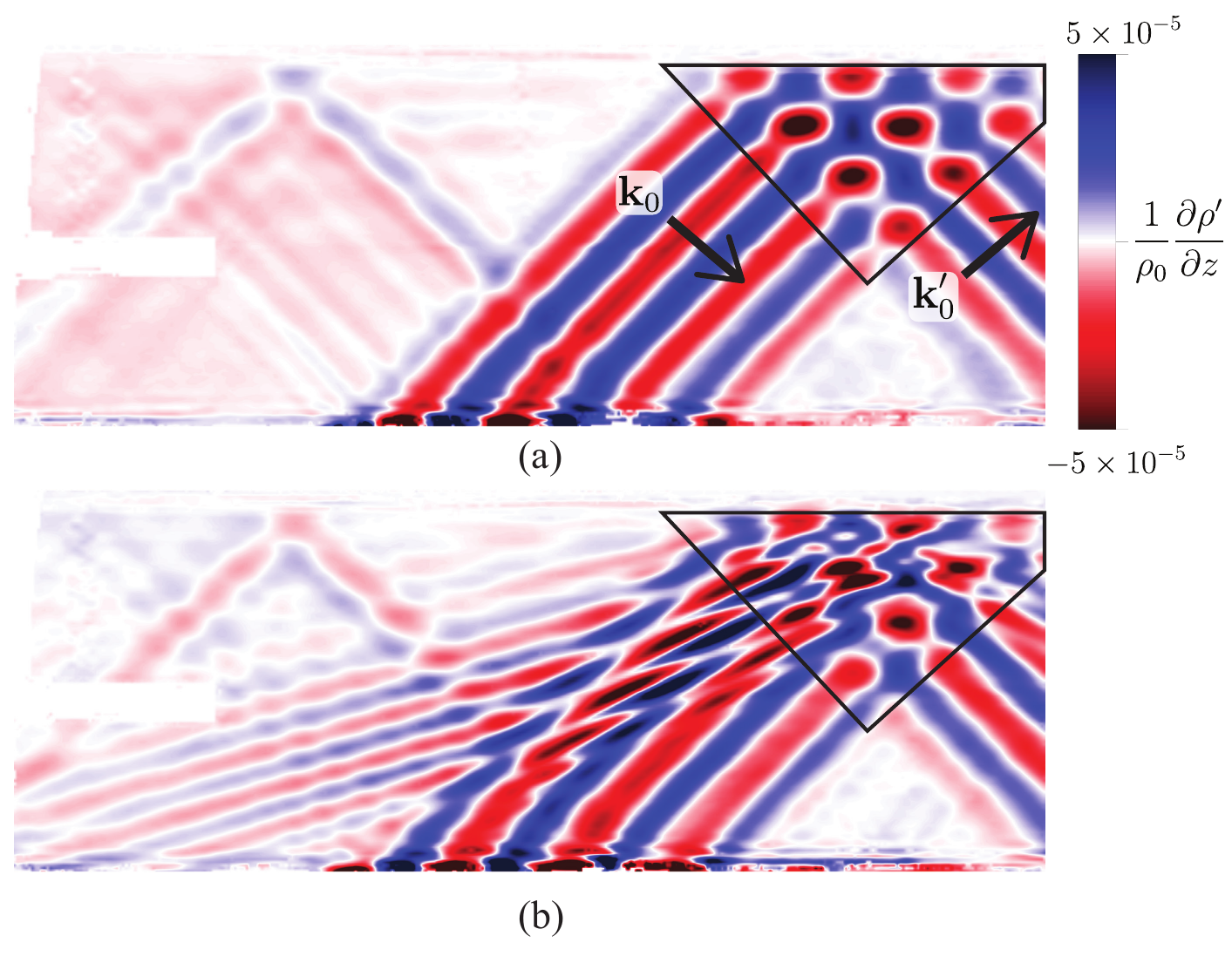}
    \caption{Examples of the reflection zone, marked by the black polygons, for (a) the primary wavebeam $\wave_0$ by itself, and when (b) secondary waves are present. The wavevectors shown illustrate the change in the vertical component that occurs through reflection. Noticeable signal propagating up and to the left, primarily from the edges of the wavebeam, is due to termination effects that are incompletely cancelled by \eqref{eqn: wave beam envelope}.}
    \label{fig:primary wave reflection zone and triad reflection zone}
\end{figure}

Figure \ref{fig:primary wave reflection zone and triad reflection zone} illustrates how reflection from the free surface introduces additional wave--wave interactions by changing the sign of the vertical component of the wavenumber vector. Within the interaction region (black outline), interaction is possible between a given wave $\wnv_j$ and its reflection that we denote $\wnv_j^\prime$, which can lead to additional interaction mechanisms that are impossible elsewhere.

% ***** Reflection zone diagram *****

\begin{figure}
    \centering
    \includegraphics[width=0.8\linewidth]{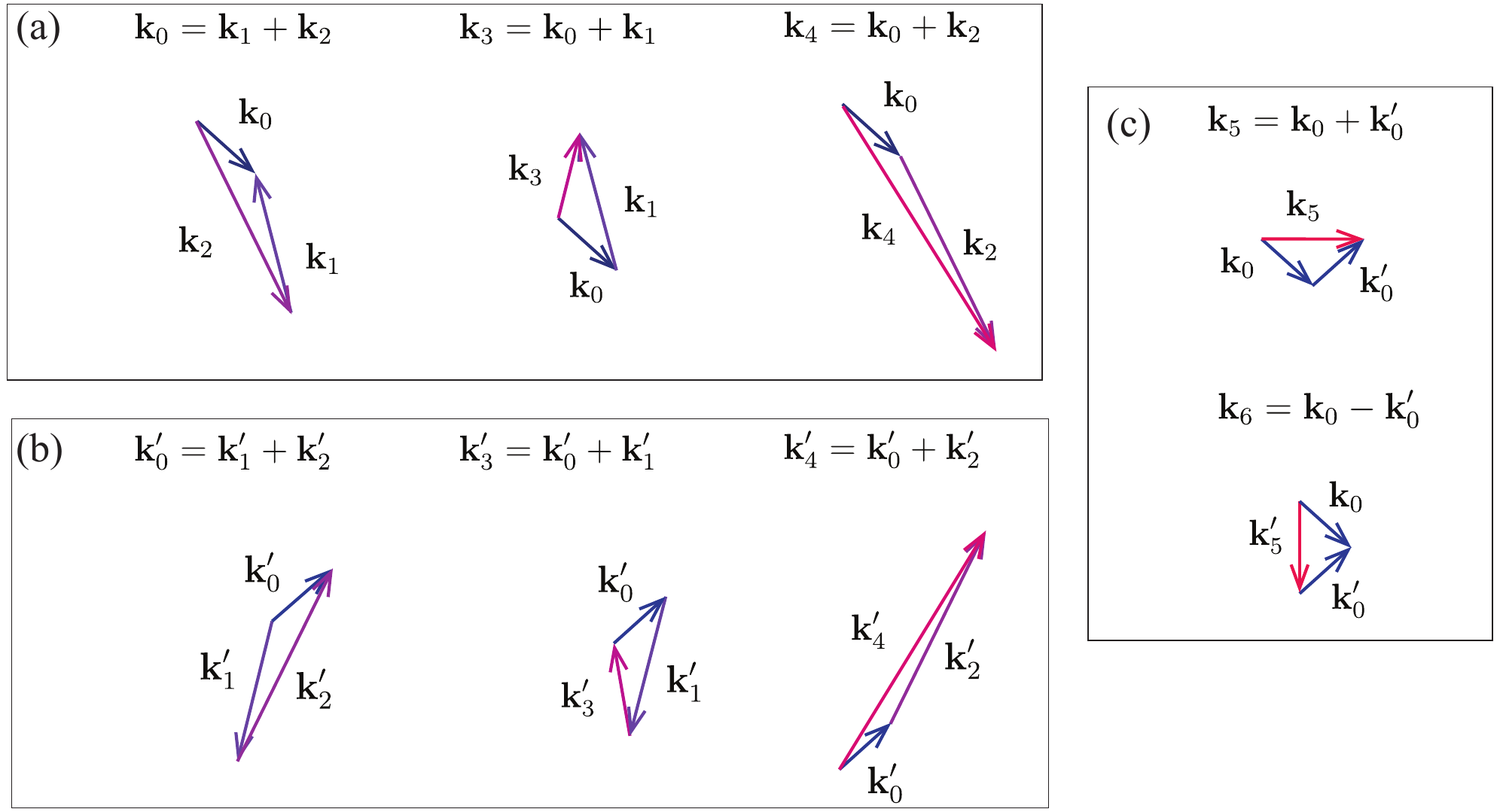}
    \caption{The first-order wave--wave combinations that can be formed for $\wnv_0$, $\wnv_3$, $\wnv_4$, and $\wnv_5$. A prime indicates a reflected wavenumber vector \ie same $x-$component, opposite $z$-component. (a) Standard wavenumbers. (b) Reflected wavenumbers. (c) Primary wave self-interactions.}
    \label{fig:first order wavenumber combinations}
\end{figure}

Figure \ref{fig:first order wavenumber combinations} illustrates candidates for standard sum interactions involving $\wave_0$ with $\wave_{1,2}$, analogous interactions involving reflected $\wave_0$ and $\wave_{1,2}$, and self-interactions of the primary wave within the reflection zone. These interactions can also generate second-order combinations involving first-order products together with $\wnv_0$ or $\wnv_0^{\prime}$. For example, $\wnv_4 = \wnv_0 + \wnv_2$ can also be written as $\wnv_4 = (\wnv_1 + \wnv_2) + \wnv_2 = 2\wnv_2 + \wnv_1$. The addition of the reflected components thus permits a significant number of additional interactions.  

% ***** Full chart of first order wave--wave interaction *****

% ***** Wavenumber diagrams matching the TRI modes *****
\begin{figure}
    \centering
    \includegraphics[width=0.8\linewidth]{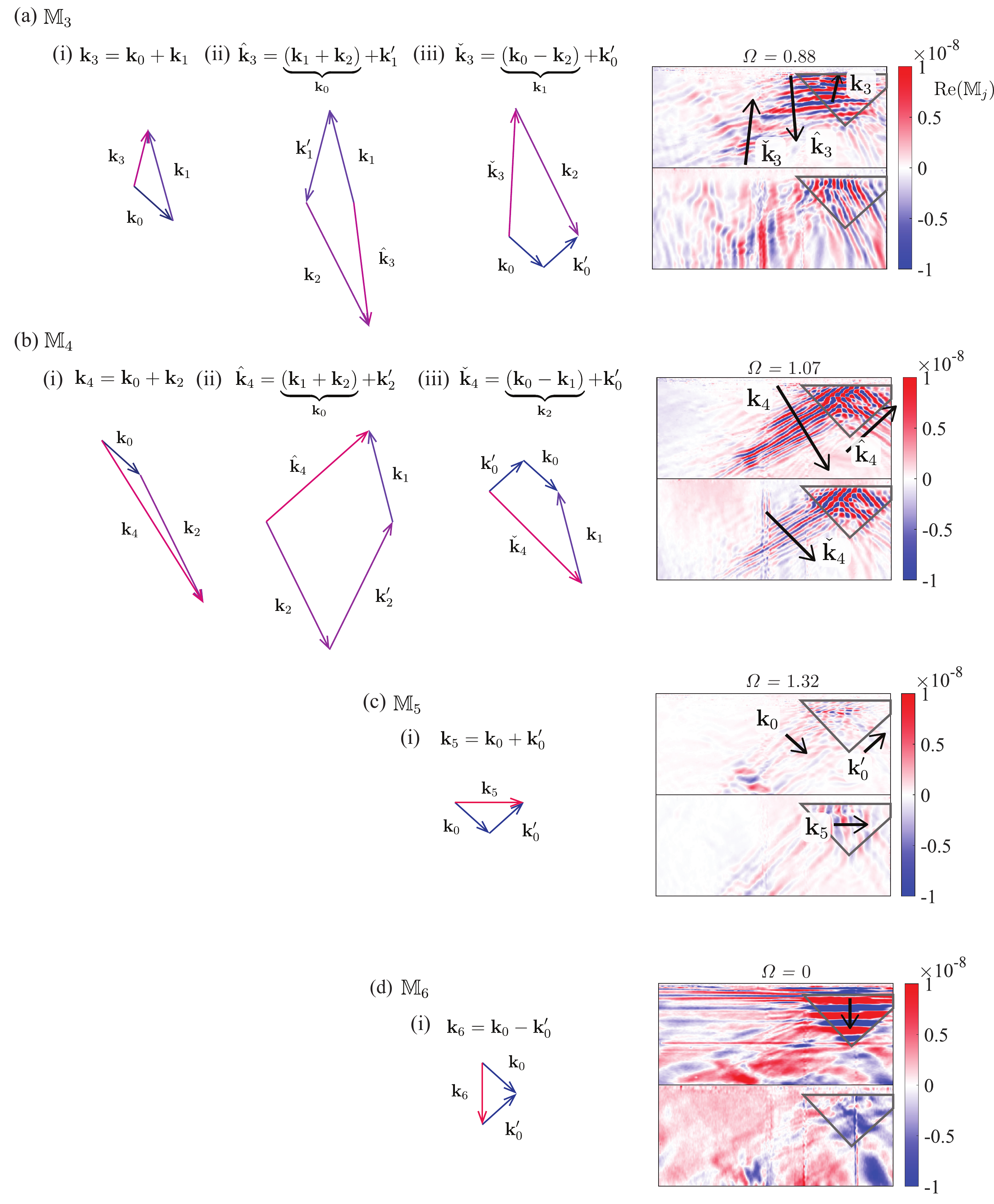}
    \caption{Wavenumber vector diagrams for some triads matching the frequencies observed in the data. The wavenumber diagrams are shown to the left, with the corresponding mode image to the right. (a) Wavenumber diagrams for frequency $\Omega_3 = \Omega_0 + \Omega_1$, with $\Omega_3 = 0.88$. (b) Wavenumber diagrams for frequency $\Omega_4 = \Omega_0 + \Omega_2$, with $\Omega_4 = 1.07$. (c) Wavenumber diagrams for frequency $\Omega_5 = 2\Omega_0$, with $\Omega_5 = 1.32$. (d) Wavenumber diagrams for frequency $\Omega_6 = 0$. The left-hand wavenumber vector diagrams are double-length versions of the right-hand ones on the mode diagrams. The reflection region is shown in the mode diagrams with a grey polygon.}
    \label{fig:wavenumber diagrams for TRI configurations}
\end{figure}

Figure \ref{fig:wavenumber diagrams for TRI configurations} illustrates the permissible first- and second-order wavenumber vector interactions for $\wave_{3,4,5,6}$, and superimposes them onto visible structure in the modes. 
For $\wave_3$ and $\wave_4$, a hat symbol over the wavenumber vector (\eg $\hat{\wnv}_r$) indicates combinations arising from summations involving a reflected wave, while a check symbol (\eg $\check{\wnv}_r$) indicates subtractions involving a reflected wave. This notation is unnecessary for $\wave_5$ and $\wave_6$, which may arise from the sum and difference of the primary wave and its reflected component (Figure \ref{fig:first order wavenumber combinations}(c)); We therefore refer to these simply as $\wnv_5$ and $\wnv_6$. 

% What the modes look like
Figure~\ref{fig:wavenumber diagrams for TRI configurations}(a) shows the $\wave_3$ mode, with frequency $\Omega_3 = \Omega_0 + \Omega_1$. Its structure is unlike that of the propagating waves $\wave_0$, $\wave_1$, and $\wave_2$: the nearly horizontal structure in the upper ($z$-gradient) part of the image is largely confined to the region where $\wave_0$ may interact with $\wave_1$, and the vertical structure in the lower ($x$-gradient) part is similarly confined. Figure~\ref{fig:wavenumber diagrams for TRI configurations}(a)(i)--(iii) shows several candidate wavenumber constructions for $\wave_3$, derived from combinations involving $\wnv_0$ and $\wnv_1$. 

% Why none of the candidates is a propagating wave
If this disturbance were a freely propagating wave, then, from the dispersion relation we would expect
\begin{equation}\label{eqn:dispersion relation}
    \cos{\theta_3} = \Omega_3,
\end{equation}
or more generally
\begin{equation}\label{eqn:dispersion relation 2 in wavenumber vector form}
    \Omega_j = \cos{\theta_j} = \frac{|\wnv^*_{j, z}|}{|\wnv_j^*|},
\end{equation}
where the subscript $z$ indicates the vertical component of the wavenumber vector and $\wnv_j^*$ is any of $\wnv_j$, $\hat{\wnv}_j$, or $\check{\wnv}_j$. From $\Omega_3 = 0.88$, \eqref{eqn:dispersion relation} gives $\theta_3 \approx 14\unit{\degree}$, the angle between the wavebeam and vertical. The candidate vectors correspond to angles of approximately $14\unit{\degree}$ for $\wnv_3$, $7\unit{\degree}$ for $\hat{\wnv}_3$, and $3\unit{\degree}$ for $\check{\wnv}_3$. The candidate wavevector that satisfies the dispersion relation, $\wnv_3 = \wnv_0 + \wnv_1$, (Figure~\ref{fig:wavenumber diagrams for TRI configurations}(a)(i)) does not match the visible structure in the mode image, while the candidates that better match parts of the observed structure (Figure~\ref{fig:wavenumber diagrams for TRI configurations}(a)(ii),(iii)) do not satisfy the corresponding temporal dispersion relation for frequency $\Omega_3$ and are not consistent with the observed wavelengths. The signal also remains confined to the interaction region, indicating that it does not propagate as a wave. We therefore conclude that $\wave_3$ is not a propagating wave, but instead a superposition of non-propagating oscillatory signals. Since $\Omega_3 = 0.88$ lies below the buoyancy frequency, this non-propagation cannot be attributed simply to frequency. Although not a wave, the $\wave_3$ oscillatory signal is retained by spDMD as a significant component required to meet the chosen reconstruction-error tolerance.

We carry out the same analysis for $\wave_4$, with the frequency of $\Omega_0 + \Omega_2= \Omega_4$, in Figure \ref{fig:wavenumber diagrams for TRI configurations}(b). Because this frequency is (very slightly) above $N$, no meaningful propagating wave angle $\theta$ can be calculated for $\wave_4$, although a three-wavevector interaction diagram can still be drawn. 
The strongest visible signal is confined within the envelope of the primary wave $\wave_0$, particularly concentrated in the interaction region close to the free surface. In this case, $\wnv_4$, $\hat{\wnv}_4$, and $\check{\wnv}_4$ all give reasonable approximations to the orientation of structures visible in parts of the observed mode (Figure \ref{fig:wavenumber diagrams for TRI configurations}(b)(i)--(iii)). As with $\wave_3$, however, 
none of these wave combinations can satisfy the dispersion relation, and $\wave_4$ is therefore best interpreted as an evanescent disturbance that decays exponentially away from the region in which it is generated.

Figure \ref{fig:wavenumber diagrams for TRI configurations}(c) shows the $\wave_5$ disturbance with $\Omega_5 = 2\Omega_0 = 1.32$. The mode structure is concentrated in the reflection zone, with fainter structure filling the area of $\wave_0$. Within the reflection zone for the top half of the image, a very fine structure is visible at an angle of approximately \SI{70}{\degree} to the vertical. This structure appears to be an interference pattern between $\wnv_5$ and $\wnv_5^{\prime}$. In contrast, the reflection region in the lower half of the image shows structure that is almost vertical and has a wavelength comparable with $\wave_0$. Since $\wnv_5$ can arise from $\wnv_0 + \wnv_0{^\prime}$, we superimpose these wavenumber vectors on Figure \ref{fig:wavenumber diagrams for TRI configurations}(c). We find $\wave_0$ and $\wave_0^{\prime}$ match well in orientation to the observed structure in the region filled by $\wave_0$, while $\wave_5$ is a good match to the vertical structure visible in the lower half of the image. 

As with $\wave_4$, the propagating wave dispersion relation cannot be satisfied for $\wnv_5$. Thus, $\wave_5$ is best interpreted as an evanescent oscillatory disturbance from a non-linear self-interaction of $\wave_0$ within the reflection zone. Examining this frequency band in Figure \ref{fig:wiggle plots, a} indicates the $\wave_5$ signal is the least robustly recovered of the additional disturbances identified by spDMD in the TRI case; the appearance of $\wave_5$ is more sporadic than $\wave_3$ and $\wave_4$. Calculation of the mode amplitudes also indicates that $\wave_5$ has the smallest amplitude of all the detected waves --- \eg Figure \ref{fig:TRI amplitude plot}: $\wave_5$ has $\frac{|\wave_j|}{|\wave_0|}$ of $\mathcal{O}(10^{-3})$ while $\wave_{1,2,3,4}$ are $\mathcal{O}(10^{-2})$ or greater --- and therefore may lie close to the threshold of being retained as a significant signal at the chosen 5\% reconstruction-error tolerance. 

Finally, Figure \ref{fig:wavenumber diagrams for TRI configurations}(d) depicts the $\wave_6$ signal with almost perfectly horizontal lines of constant phase that appear only within the reflection zone. This matches well with ${\wnv}_6 = \wnv_0 - \wnv_0^{\prime}$ shown in Figure \ref{fig:wavenumber diagrams for TRI configurations}(d), which corresponds to a vertical wavenumber vector, horizontal lines of constant phase, and a frequency that also satisfies the dispersion relation. Thus, $\wave_6$ has some of the properties of a propagating wave, but appears confined to the reflection zone, and it therefore better characterised as a confined steady osclllating signal rather than as an evanescent or freely propagating wave. Because contours of phase angle do not propagate away from the reflection zone, it is not possible to attribute a group velocity, so the sign of the wavevector is essentially arbitrary. 

We therefore conclude that the observed $\wave_{3,4,5}$ are all non-propagating oscillatory signals arising from additional wave--wave interactions in the flow, and driven by energy from the triad of $\wave_0$ and $\wave_{1,2}$. As noted in \S\,\ref{subsec:TRI} to \S\,\ref{subsec:TTR}, both the secondary waves $\wave_{1,2}$ and the higher-frequency signals $\wave_{3,4,5}$ exhibit frequency and amplitude fluctuations. While the observation of frequency fluctuation in the secondary waves is not novel, being first noted by \citet{Grayson2022}, we believe the extended observations of fluctuation in the higher frequencies have not been previously reported. We summarise the characteristics of the three response types (and the no-response regime) in Table \ref{tab:table of response types and their components}. 

The higher-frequency signals are detected more intermittently in STR than in TRI, but we attribute this primarily to amplitudes falling intermittently below the significance threshold for a given reconstruction-error tolerance, rather than to their intermittent non-existence. Since TRI occurs at higher forcing amplitudes (Table~ \ref{tab:table of wave amplitudes for each regime}), it can supply more power to the secondary and tertiary signals, making them easier to retain at the same reconstruction-error tolerance. Tighter error tolerances are not necessarily beneficial: below $5\%$, additional retained modes often have amplitudes $\frac{|\wave_j|}{|\wave_0|} = \mathcal{O}(10^{-3})$ or smaller and are indistinguishable from experimental noise, thereby polluting the clarity of the recovered mode set. 

\begin{table}[h]
    \centering
    \begin{tabularx}{0.9\linewidth}{X X X X}
        \toprule
        Response & Wave components & Oscillatory signals & Steady disturbance\\
        \midrule
         TRI & $\wave_0$, $\wave_1$, $\wave_2$ & $\wave_3$, $\wave_4$, $\wave_5$ & $\wave_6$ \\
         STR &  $\wave_0$, $\wave_1$, $\wave_2$ & $\wave_3$, $\wave_4$, $\wave_5$ & $\wave_6$ \\
         TTR & $\wave_0$, $\wave_1$, $\wave_2$ & & \\
         No response & $\wave_0$ &  & \\
         \bottomrule
    \end{tabularx}
    \caption{Detected components of each type of triadic response.}
    \label{tab:table of response types and their components}
\end{table}

\subsection{Statistical results across all experiments}\label{sec:comparison of results}

\begin{figure}
    \centering
    \includegraphics[width=12.9 cm]{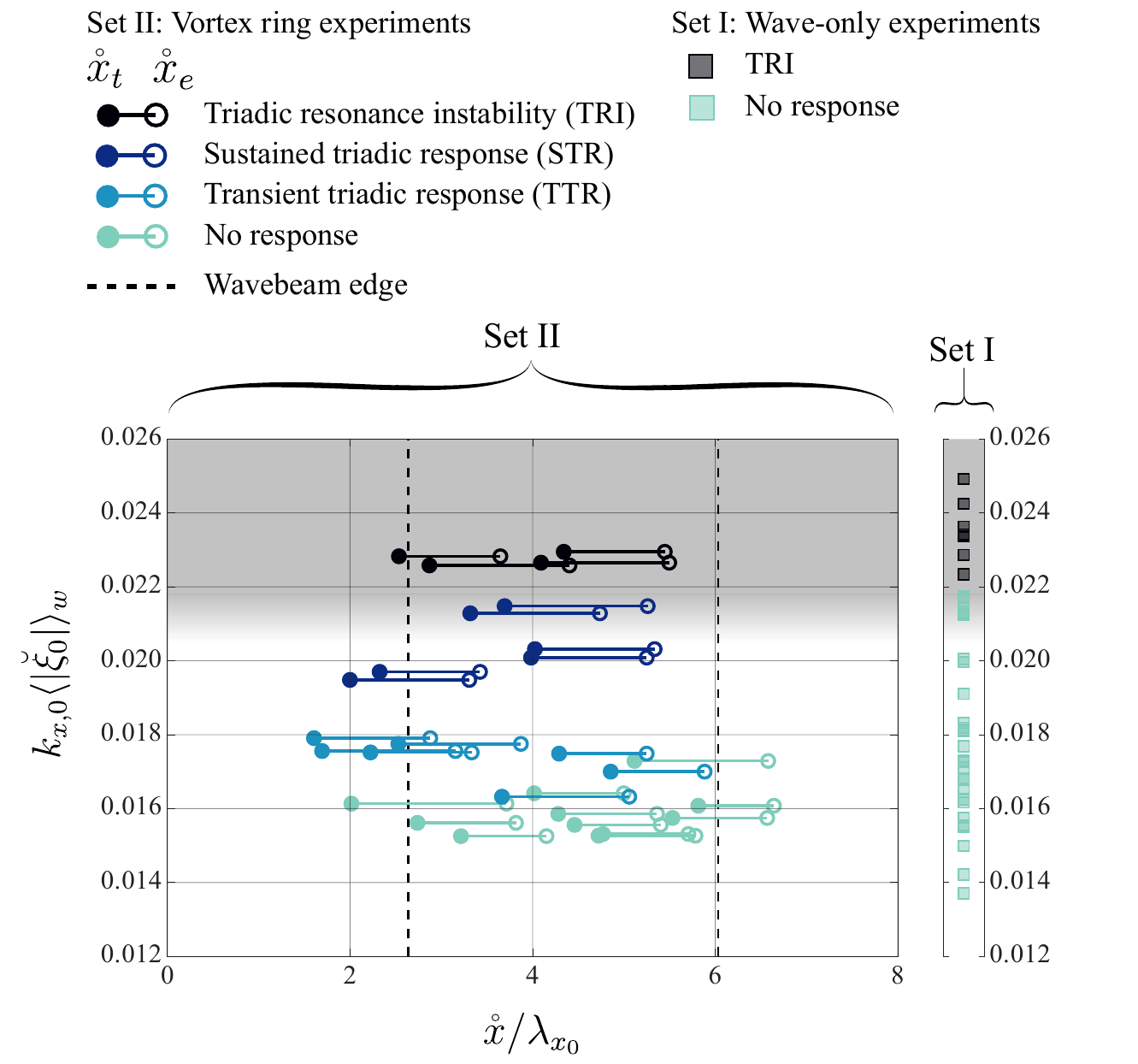}
    \caption{A summary of results from full set of experiments. The main panel shows the Set \setb experiments containing the interaction of $\wave_0$ with a vortex ring. The transition ($\mathring{x}_t$) and end point ($\mathring{x}_e$) distances of vortex rings fired into a wave beam are plotted against the spatially-averaged amplitude of $\wave_0$ for each experiment. A horizontal line links $\mathring{x}_t$ and $\mathring{x}_e$ for each experiment. The vertical dashed ines indicate the lateral extent of the wavebeam where the axis of the vortex tube intersects the beam. Colour indicates the wavebeam response to the vortex ring, with dashed black vertical lines showing width of $\wave_0$ at the height of the vortex rings. The slim vertical panel on the right shows the results from Set \setw, which illustrates the amplitude threshold above which linear TRI occurs, shown in grey shading for both panels.}
    \label{fig:full set of results} 
\end{figure}

We now summarise the results from all 64 experiments analysed using the chosen $5\%$ reconstruction-error tolerance (Figure~\ref{fig:full set of results}). The 35 wave-only experiments in Set \setw are shown in the right-hand panel and define the lower bound for the spontaneous onset of TRI in the absence of external perturbation. From these experiments we estimate the TRI threshold to be ${\pwx}\langle |\breve{\xi}_0| \rangle_w \approx 0.0218$, with an uncertainty band over $0.0205 \lesssim {\pwx}\langle |\breve{\xi}_0| \rangle_w \lesssim 0.0218$ to account for experimental variability. 

The 29 vortex-ring experiments in Set \setb are shown in the main left-hand panel. These span amplitudes both above and below the linear TRI threshold and show how a vortex-ring perturbation can lead to four outcomes: above-threshold TRI, below-threshold STR, below-threshold TTR, and no response. At the lowest amplitudes, the perturbation does not trigger any discernible triadic state; these cases are classified as having no response in Tables~\ref{tab:table of wave amplitudes for each regime} and~\ref{tab:table of response types and their components}. 

While our spDMD analysis has revealed substantial new richness in an existing data set, it is nevertheless reassuring that it leads to the same triadic state classifications reported in \citet{Grayson2024} using a standard fixed-rank approach to DMD with 9 singular values. We note that while possible, to obtain equivalent observations to the present work using fixed-rank DMD would have required laborious manual tuning individually for each DMD window. With this would come a corresponding loss of objectivity, especially if signals observed in an earlier DMD window were actively sought in a later time window.

% ***** Main results plot, with  vortex rings in main panel *****
% ****************************************************************

%%%%%%%%%%%%%%%%%%%%%%%%%%%%%%%%%%%%%%%%%%%%%%%%%%%%%%%%%%%%%%%%%%%%%
\section{Discussion}\label{sec:Discussion}

\subsection{Relationship to prior work on wave attractors and surface waves}\label{subsec:Relationship to prior work on wave attractors and surface waves}

Previous experimental observations of wave--wave interactions following TRI \citep{Scolan2013, Brouzet2016b, Brouzet2017, Davis2020} have largely been in the context of internal wave attractors, where geometry focuses a primary wave until its amplitude exceeds the linear threshold for TRI \citep{Scolan2013, Brouzet2016b, Brouzet2017} or, alternatively, where the forcing amplitude is gradually increased during the experiment \citep{Davis2020}.

\citet{Scolan2013} observed TRI in a trapezoidal domain, reporting additional spectral peaks at $\Omega_3 = 0.85$ and $\Omega_4 = 1.0$ attributed to to $N \cos{\mu}$ and $N$, respectively, where $\mu$ is the sloping wall angle. 
These values are close to $\Omega_0 + \Omega_1$ and $\Omega_0 + \Omega_2$ from our system, suggesting similar sum interactions were present. The relative magnitudes were also consistent, with higher-frequency waves having lower magnitudes: $|\wave_0| \sim \mathcal{O}(10^{0})$, with $|\wave_{1,2}|\sim\mathcal{O}(10^{-1})$ and $|\wave_{3,4}|\sim\mathcal{O}(10^{-2})$. However, as both \citet{Scolan2013} and \citet{Brouzet2016b} --- who used the same geometry and forcing frequency --- focused their analysis on the $\Omega_{0,1,2}$ triad, it remains unclear whether $\Omega_{3,4}$ corresponded to propagating waves.

\citet{Brouzet2017} observed a cascade of sum and difference interactions between secondary waves arising from TRI. Beginning with the difference $\Omega_{-} = \Omega_1 - \Omega_2$, this combines with the primary wave to produce 
$\Omega_{+} = \Omega_0 - \Omega_{-}$, and the cascade extends further, generating a countably infinite discrete frequency set. Beam focusing in the attractor geometry makes these higher-order signals more visible and long-lasting than in non-focusing configurations. In our system, the equivalent first step would produce $\Omega_2 - \Omega_1 = \Omega_{-} \approx 0.16$ and $\Omega_0 - \Omega_{-} = \Omega_{+} \approx 0.50$, but these are absent from our analysis, likely due to their association with side-wall and slope reflections. Crucially, all interactions in \citet{Brouzet2017} occur below $N$ and satisfy both dispersion and resonance conditions, whereas our higher-frequency signals --- with the exception of $\wave_3$ and $\wave_6$ --- live above $N$ and do not satisfy the internal wave dispersion relation. 

\citet{Davis2020} observed the same sum-frequency structure of $\Omega_3 = \Omega_0 + \Omega_1$ and $\Omega_4 = \Omega_0 + \Omega_2$, extending this pattern to $3\Omega_0$ and explicitly showing above-$N$ instances. However, as their study focused on wave turbulence, the properties of these higher-frequency signals were not characterised. The frequency and amplitude fluctuations we observe are also absent from their time-frequency diagrams, though this may reflect their use of a long FFT window relative to our fluctuation periods. 

\citet{Rodda2022} identified internal bound waves in TRI experiments in a pentagonal domain. These are wave-like structures that need not satisfy the dispersion relation but can nevertheless participate in non-resonant transfers of power. Notably, their bound wave signatures remained visible above $N$, demonstrating that such signals, which we expect also to be present in the ocean,   can still contribute to the energy spectrum at frequencies greater than $N$. While we do not observe instances of bound waves --- which would manifest as additional spatial structure at frequencies close to $\Omega_{1,2}$ --- the implication for power transfer is directly relevant: the high-frequency oscillatory disturbances we observe may similarly participate in power transfer at frequencies where internal wave propagation is prohibited. 

The oscillatory signals reported here have clear analogues in other types of TRI experiments, appearing either as propagating waves below the buoyancy frequency, or as measured but uncharacterised signals. Our characterisation, however, is new, as is the observation of such signals in below-threshold triadic states such as STR.

\subsection{Relationship to previous work on the Magic Carpet}\label{subsec:Relationship to previous work on the Magic Carpet}

The $\wave_{3,4,5}$ oscillatory signals identified by spDMD appear to be generated by interaction between two already-existing waves, rather than by a triadic resonance instability in which one already-existing wave gives rise to two secondary waves. This distinction motivates a comparison with \citet{Dobra2022}, who systematically explored cascades of two-wave interactions using the Magic Carpet. An open question is whether these oscillatory signals are coupled to the quasi-periodic frequency fluctuations we observe in $\wave_{1,2, 3, 4}$, corroborating earlier observations of fluctuation in $\Omega_{1,2}$ by \citet{Grayson2022}.

\citet{Dobra2022} explored interactions of two internal wavebeams using the Magic Carpet, carrying out a hierarchical decomposition into first, second, and third-order interactions. Applying their framework to our $\Omega_0 + \Omega_1 = \Omega_3$ interaction yields
\begin{equation*}
    \Omega_0, \quad  \Omega_1, \quad \underbrace{\Omega_0 + \Omega_1}_{\Omega_3}, \quad \text{and} \quad  \underbrace{\Omega_0- \Omega_1}_{\Omega_2}, 
\end{equation*} where our full TRI triad (with $\Omega_2$) emerges at second-order from a single `branch' (\ie from a single $\Omega_0 \otimes \Omega_1 \to\Omega_3$ interaction, where $\otimes$ indicates interaction between). Unlike \citet{Dobra2022}'s two-wave system, however, TRI produces two secondary waves, meaning each branch couples to the other: third-order interactions from the $\Omega_0 \otimes \Omega_1$ branch contribute frequency components into $\Omega_2$ and $\Omega_4$, which are the frequencies belonging to the $\Omega_0 \otimes \Omega_2$ branch, and vice versa. This cross-branch coupling suggests the possibility of frequency-domain feedback between all five components $\Omega_{0,1,2,3,4}$, with higher-order terms such as $\Omega_0 + \Omega_3$ not observed in our spDMD, consistent with their expected lower amplitude.

\citet{Dobra2022} found that the second-order difference component ($\Omega_b - \Omega_a$) was significantly larger in amplitude than the sum ($\Omega_b + \Omega_a$) component, by approximately an order of magnitude. The equivalent in our system is $\Omega_0-\Omega_1 = \Omega_2$ being significantly larger in amplitude than $\Omega_0 + \Omega_1 = \Omega_3$, which is what we observe for both TRI and STR. In the TTR case, where $\Omega_3$ is absent and thus contributes no second or third-order terms back into $\Omega_0$ or $\Omega_1$, $\Omega_2$ is indeed of lower amplitude, though we cannot disentangle this from the lower forcing amplitude of the primary wave in TTR. 

The cross-branch coupling identified above offers a potential mechanism for the negatively-correlated frequency fluctuations between $\wave_{1,3}$ and $\wave_{2,4}$ observed in TRI and STR. When the frequency of one branch (\eg $\Omega_0 \otimes \Omega_1 \to\Omega_3$) increases, its third-order contributions feed into $\Omega_0$, $\Omega_2$, and $\Omega_4$ (the frequencies of the other branch) potentially leading to correlated changes there. This feedback mechanism may thus explain the fluctuations we observe. TTR cases, where $\wave_{3,4}$ are absent or very weak, offer a test case: if, for example, longer-lasting TTR experiments entirely lack the quasi-periodic fluctuations seen in TRI and STR, this would strongly suggest that the cross-branch feedback drives these fluctuations. 

The oceanographic relevance of these findings may be limited by factors including reflections from 3D topography, Doppler shifting due to mean currents, and non-linear stratifications. Nevertheless, our specific configuration is directly relevant to the case of wave impingement on the underside of a pycnocline, and our characterisation of oscillatory signals generated by TRI and STR and their potential role in redistributions at frequencies greater than $N$ adds to the complex physical picture of how energy is distributed in stratified wave fields such as the ocean.

\section{Conclusions}\label{sec:Conclusions}

Through spDMD analysis, we have identified previously uncharacterised oscillatory disturbances in TRI and STR that are locally confined, do not satisfy the internal wave dispersion relation, and are absent in TTR. Following the hierarchical wave--wave decomposition approach of \citet{Dobra2022}, we find these signals arise from interactions between the primary and secondary waves, with cross-branch coupling between interaction pairs offering a potential mechanism for the long-period frequency and amplitude fluctuations observed both here and in \citet{Grayson2022} and \citet{Grayson2024}. The evidence for cross-branch coupling as a driver for the frequency fluctuations is convincing, but its role in amplitude variation is less certain. Future work with longer-duration TTR, in which these oscillatory disturbances are absent, would provide a direct test of whether this cross-branch feedback is the origin of the quasi-periodic fluctuations observed in TRI and STR. 

% Specify following sections are appendices. Use \appendix* if there
% only one appendix.
\FloatBarrier
%%%%%%%%%%%%%%%%%%%%%%%%%%%%%%%%%%%%%%%%%%%%%%%%%%%%%%%%%%%%%%%%%%%%%%%%%%%%%%%
\appendix

\FloatBarrier
%%%%%%%%%% Appendix 1 %%%%%%%%%%%%%%%%%%%%%%%%%%%%%%%%%%%%%%%%%%%%%%%%%%%%%%%%%
\section{Sparsity-promoting DMD: matrices and notation}\label{appendix:spdmd matrices}

We present here the notation used for our implementation of \citet{Jovanovic2014}'s spDMD, as well as for the matrices of standard DMD. The standard DMD algorithm begins with a state vector $x_j$, which represents one frame from a sequence of flow field measurements. We denote a sequence of $N$ snapshots of a given field as
\begin{equation}
    \underbrace{X_0}_{p\times N} = 
    \begin{bNiceMatrix}[last-row,nullify-dots]
        \mid & \mid & \mid &  & \mid & \mid \\
        x_0 & x_1 & x_2 & \cdots & x_{N-2} & x_{N-1} \\ 
        \mid & \mid & \mid &  & \mid & \mid \\
        \underbrace{}_{p\times 1} & & & & &  
    \end{bNiceMatrix},
\end{equation} where an equal time interval of $\Delta t$ is assumed between two subsequent snapshots $x_j$ and $x_{j+1}$ and each element in $x_j$ corresponds to one of $p = m \times n$ pixels in the original image. DMD defines the evolution of the matrix of state vectors as
\begin{align}\label{eqn:dmd matrices}
    {\underbrace{\vphantom{AX_{j+1}}{X_1}}_{p \times N}} = {\underbrace{\vphantom{AX_{j+1}}{A}}_{p \times p}}{\underbrace{\vphantom{AX_{j+1}}{X_0}}_{p \times N}}
\end{align}
where
\begin{equation}
    \underbrace{X_1}_{p\times N} = 
    \begin{bNiceMatrix}[last-row,nullify-dots]
        \mid & \mid & \mid &  & \mid & \mid \\
        x_1 & x_2 & x_3 & \cdots & x_{N-1} & x_{N} \\ 
        \mid & \mid & \mid &  & \mid & \mid \\
        \underbrace{}_{p\times 1} & & & & &   
    \end{bNiceMatrix}
\end{equation} is the new set of state vectors created from advancing $X_0$ by $\Delta t$ and $A$ is a linear mapping linking the matrix of flow fields $X_0$ to $X_1$. Rank truncation is generally implemented via a threshold on the singular values to produce a more compact representation. Full details, including the singular value decomposition, are given in \citet{Schmid2010} and \citet{Tu2014}. 

For spDMD, DMD is applied to the dataset with no rank truncation. Via singular value decomposition, which decomposes the snapshot matrix as $X_0 = U\Sigma V^T$, DMD yields the eigenvalue matrix
\begin{equation} 
V = 
    \begin{bNiceMatrix}
    v_1 &  & & \\
     & v_2 &  &  \\
     & & \ddots & \\
    & & & v_N \\
    \end{bNiceMatrix},
\end{equation} and the eigenvector matrix  
\begin{equation}
    W =     
    \begin{bNiceMatrix}
        w_1 & w_2 & \cdots & w_{N}.  \\ 
    \end{bNiceMatrix}
\end{equation} The DMD modes $\phi_j$ are obtained from the columns of $W$, via 
\begin{equation}\label{eqn:dmd, exact dmd modes Phi}
    \Phi = X_1 V \Sigma ^{-1}W.
\end{equation}

The spDMD then casts the data sequences $X_0$ as being composed of columns $\chi_j$, given by
\begin{equation}
    \underbrace{[\chi_0 \cdots \chi_{N-1}]}_{X_0} \approx \underbrace{[\phi_1 \cdots \phi_{N}]}_{\Phi}\underbrace{ 
    \begin{bNiceMatrix}
    \;\alpha_1 & & \\
    & \ddots & \\
    & & \;\alpha_{N}
    \end{bNiceMatrix}}_{\text{Diag}(\alpha)}
    \underbrace{ 
    \begin{bNiceMatrix}
    1 &v_1 & \cdots & v_1^{N} \\
     \vdots & \vdots & \ddots & \vdots\\
    1 & v_{N-1} & \cdots & v_{N}^{N-1}
    \end{bNiceMatrix}}_{V_\text{and}},
\end{equation} where $\phi$ are the standard DMD modes, $\alpha$ are the mode amplitudes, and $V_\text{and}$ is a Vandermonde matrix comprised of powers of the eigenvalues $v_j$. 

As noted by \citet{Schmid2022}, a naive approach to mode selection at this point, by choosing the modes with the highest amplitudes, may fail when applied to data with statistical outliers or high-amplitude noise, which is common for experimental data. In such cases, noise may manifest as structures with high amplitude but large decay rates, which appear significant but are irrelevant to the overall flow. The problem is made worse if too many singular values are retained, leading to over-fitting to some aspects of the signal. 

The spDMD method then uses the alternating direction method of multipliers to solve the optimisation problem
\begin{equation}\label{eqn:spdmd objective function with penalty term}
    \alpha_{opt} = \min_{\alpha}\,J(\alpha)
\end{equation} where 
\begin{equation}
     J(\alpha) = \lVert X_0 - \Phi\,\text{Diag}(\alpha)\,V_\text{and}\rVert_F + \gamma\lVert{\alpha}\rVert_1,
\end{equation} $\alpha_{opt}$ is the optimal amplitude vector and $\gamma$ is a positive user-defined parameter that acts as a Lagrange multiplier balancing the relative significance of sparsity versus reconstruction error. 

%%%%%%%% Appendix 2 %%%%%%%%%%%%%%%%%%%%%%%%%%%%%%%%%%%%%%%%%%%%%%%%%%%%%%%%%%%%%%%%%%%%%%%%%%%%%%%
\section{Sensitivity analysis on transient triadic responses}\label{Appendix:sensitivity on TTR}

\begin{figure}
    \centering
    \includegraphics[width=0.8\linewidth]{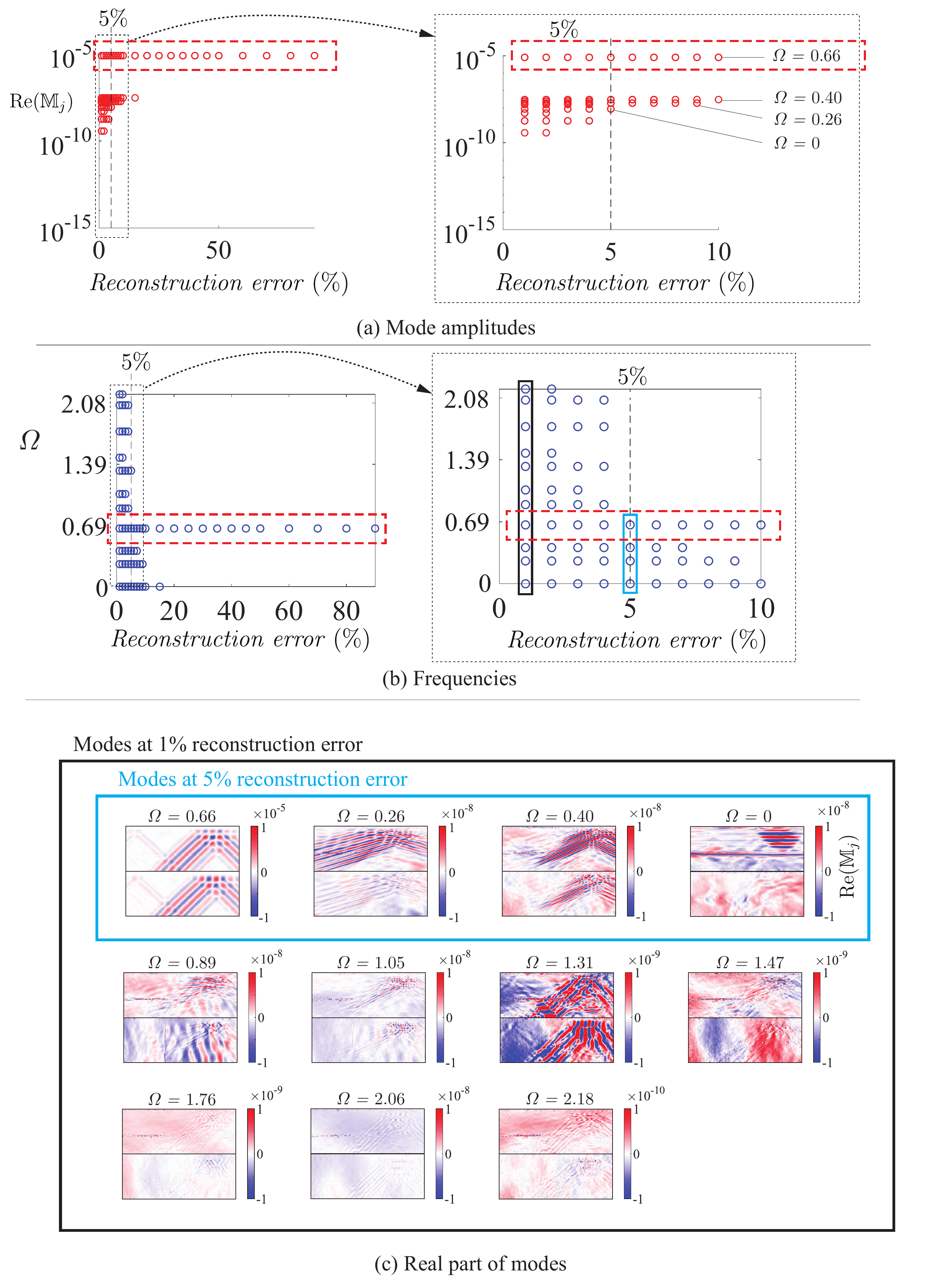}
    \caption{Analysis of mode selection by spDMD algorithm on case of TTR, for reconstruction error limits between 1\% and 90\%. Dashed red lines indicate the forcing frequency $\omega_0$.}
    \label{fig:TTR sensitivity check}
\end{figure}

Here, we present an example of an example of a sensitivity analysis carried out on a transient or decaying triadic case (TTR). Similar to Figure \ref{fig:TRI sensitivity check}, Figure \ref{fig:TTR sensitivity check} shows how changing the reconstruction error limit influences the frequency and mode magnitudes that arise from the data. 

With a 5\% reconstruction error limit, the spDMD selected fewer modes than in the TRI example (Figure \ref{fig:TRI sensitivity check}) and these identified primarily the expected TRI triad of $\Omega_0 = \Omega_1 + \Omega_2$ (Figure \ref{fig:TTR sensitivity check}c) and have a comparable order of magnitude to their equivalents in the standard case of triadic resonant instability. Notably, at this reconstruction error limit, spDMD did not select additional higher frequency modes that were selected in the standard triadic resonant instability example of Figure \ref{fig:TRI sensitivity check}; however, these modes were included when the limit was tightened to 1\%. The real part of these additional modes showed much weaker structure compared to the TRI equivalents, indicating they are of lesser significance.  

% ***** TTR sensitivity example *****
% ****************************************

%%%%%%%%%%%%%%%%%% Appendix 3: Modes %%%%%%%%%%%%%%%%%%%%%%%%%%%%%%%%%%%%%%%%%%%%%%%%%%%%%%
\FloatBarrier
\section{Modes from sustained triadic response and transient triadic response}\label{Appendix: STR and TTR modes}

\begin{figure}
    \centering
    \includegraphics[width=1\linewidth]{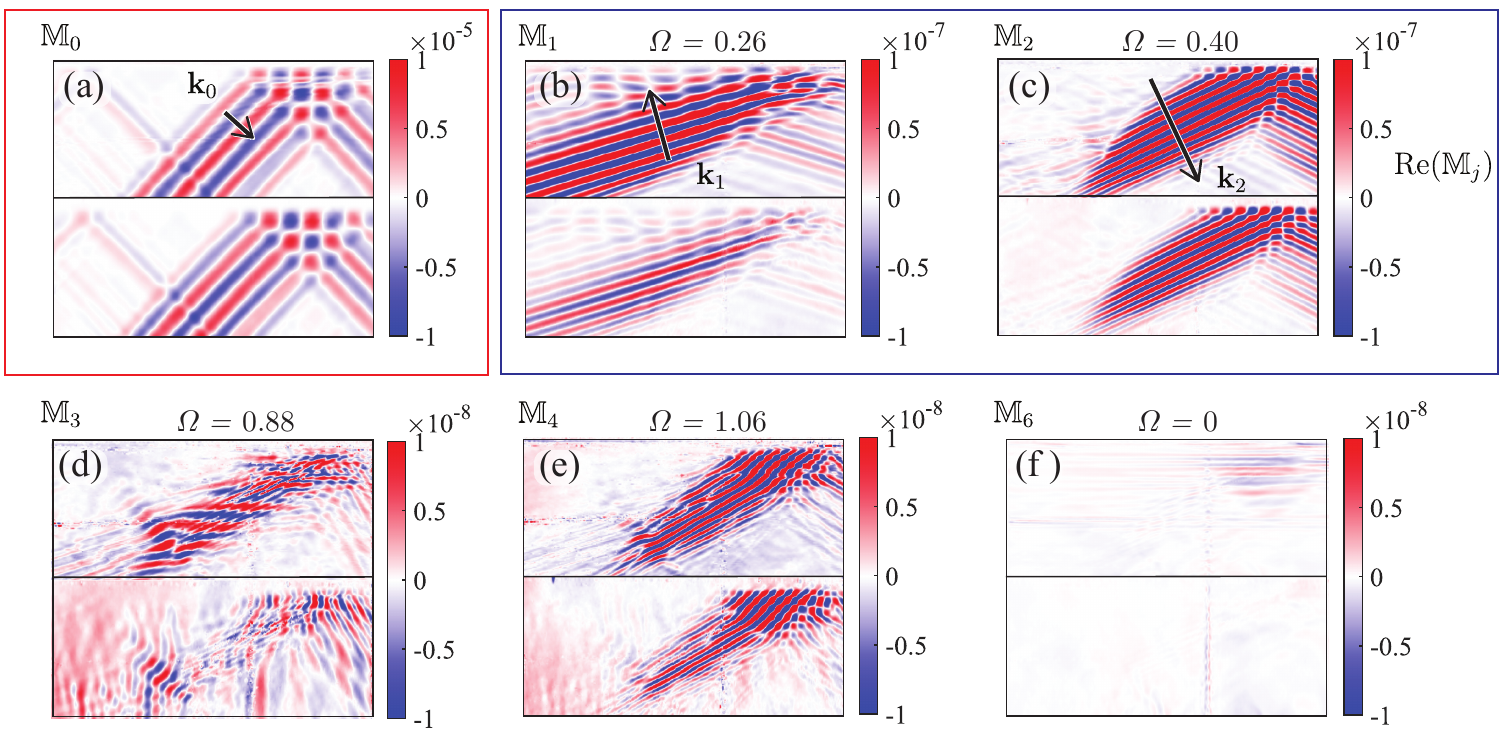}
    \caption{Modes corresponding to the frequencies from below-threshold $(\wnxa{0}\langle |\breve{\xi}_0|\rangle_w\approx 0.0195)$ STR case shown in Figure \ref{fig:wiggle plots, b}. The frequencies in order from left to right, top to bottom, are $\wave_0$: $\Omega = 0.66$, $\wave_1$: $\Omega = 0.26$, $\wave_2$: $\Omega=0.40$, $\wave_3$: $\Omega = 0.88$, $\wave_4$: $\Omega = 1.06$, and $ \wave_6$: $\Omega=0$. The modes are from the interval centred on $\tau = 169$, with the colour bar for each image indicating the amplitude of the real part of the spDMD mode. Note that the scales for the colour bars differ between images. The primary wave is boxed in red, with the two secondary waves that form the $\Omega_0 = \Omega_1 + \Omega_2$ triad boxed in blue. Arrows drawn on the images are calculated and scaled from the wavenumbers, based on the wavelength in the image.}
    \label{fig:vortex STR modes}
\end{figure}

To avoid repeating relatively similar images in the main text, modes from the STR and TTR case studies in \S\,\ref{subsec:characterisation of regimes} are instead presented here. Figure \ref{fig:vortex STR modes} shows an example of STR modes from an experiment with $\wnxa{0}\langle |\breve{\xi}_0|\rangle_w\approx 0.0195$. Comparing these modes with the TRI modes in Figure \ref{fig:TRI modes} reveals the STR modes have nearly identical structure and frequencies. 

\begin{figure}
    \centering
    \includegraphics[width=0.9\linewidth]{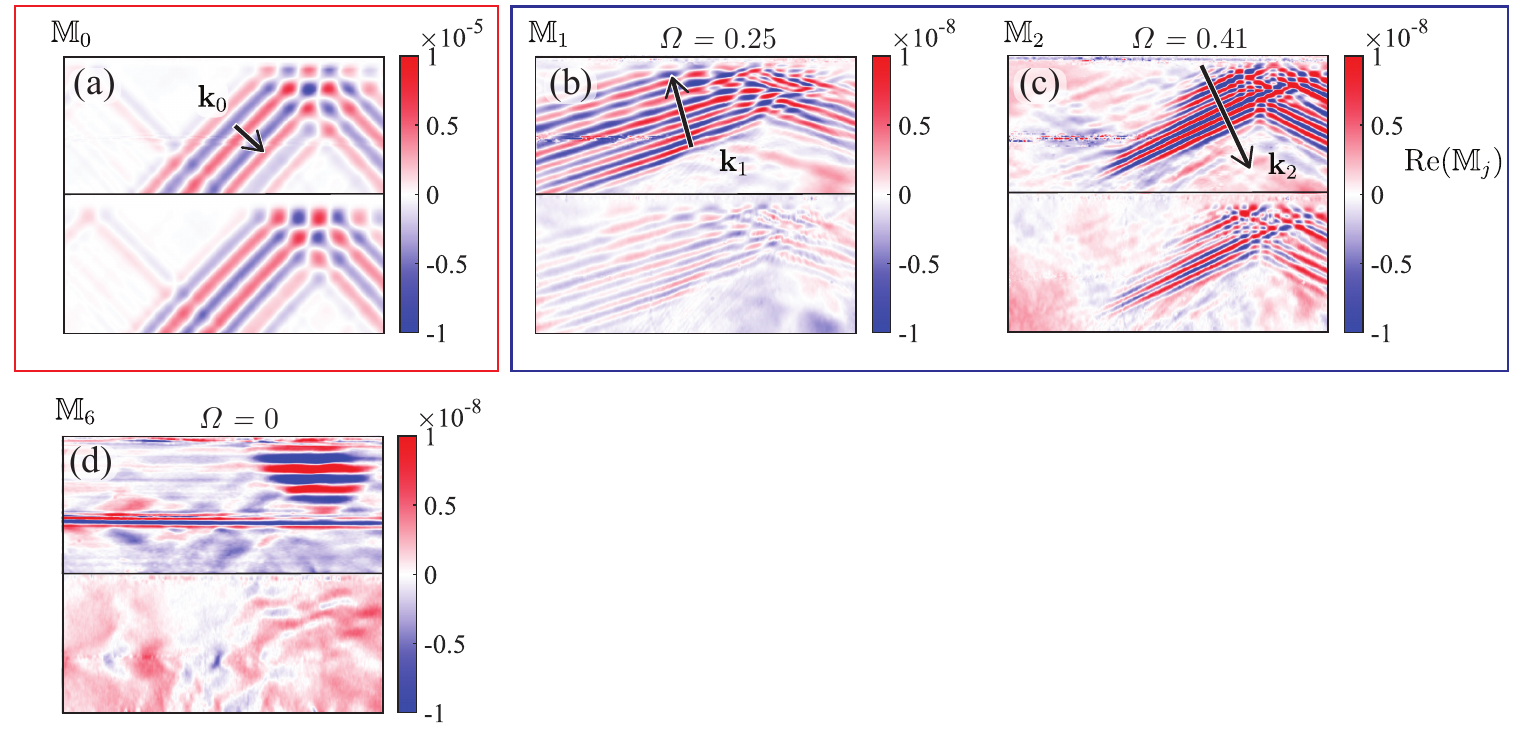}
    \caption{Modes corresponding to the frequencies from below-threshold $(\wnxa{0}\langle |\breve{\xi}_0|\rangle_w\approx 0.0170)$ TTR case shown in Figure \ref{fig:wiggle plots, c}. The frequencies in order from left to right, top to bottom, are $\wave_0$: $\Omega = 0.66$, $\wave_1$: $\Omega = 0.25$, $\wave_2$: $\Omega = 0.41$, and $\wave_6$: $\Omega = 0$. The modes are from the interval centred on $\tau = 136$, with the colour bar for each image indicating the amplitude of the real part of the spDMD mode. Note that the scales for the colour bars differ between images. The primary wave is boxed in red, with the two secondary waves that form the $\Omega_0 = \Omega_1 + \Omega_2$ triad boxed in blue. Arrows drawn on the images are calculated and scaled from the wavenumbers, based on the wavelength in the image.}
    \label{fig:TTR modes}
\end{figure}

Figure \ref{fig:TTR modes} shows an example of TTR modes from an experiment with $\wnxa{0}\langle |\breve{\xi}_0|\rangle_w\approx 0.0170$. Again referring back to Figure \ref{fig:TRI modes}, we see the lack of oscillatory signals $\wave_{3,4,5}$ or disturbance $\wave_6$ in the TTR case, as well as the clear TRI triad of $\wave_{0,1,2}$ waves.

% ***** STR modes *****

%  TTR modes plot

\FloatBarrier

% If you have acknowledgments, this puts in the proper section head.
\begin{acknowledgments}
L. J. Irvine acknowledges the support from the Defense Science and Technology Laboratories (DSTL) under grant no. DSTLX-1000160304. 

\end{acknowledgments}

% Create the reference section using BibTeX:
\bibliography{references}

\end{document}